\documentclass[sigconf]{acmart}

\usepackage{booktabs}

\usepackage{graphics}
\usepackage{subfig}
\usepackage{comment}
\usepackage{multirow}
\usepackage{xcolor}
\usepackage[linesnumbered,ruled,vlined]{algorithm2e}

\SetCommentSty{mycommfont}
\usepackage{url}

\setcopyright{rightsretained}

\acmDOI{xxxxxxx}

\acmISBN{xxxxxxxxxxxxxxx}

\begin{document}
\title{From Appearance to Essence: Comparing Truth Discovery Methods without Using Ground Truth}

\author{Xiu Susie Fang}
\affiliation{
  \institution{Department of Computing, Macquarie University}
  \city{Sydney} 
  \state{NSW} 
  \postcode{2109}
  \country{Australia}
}
\email{xiu.fang@students.mq.edu.au}

\author{Quan Z. Sheng}
\affiliation{
  \institution{Department of Computing, Macquarie University}
  \city{Sydney} 
  \state{NSW} 
  \postcode{2109}
   \country{Australia}
}
\email{michael.sheng@mq.edu.au}

\author{Xianzhi Wang}
\affiliation{
  \institution{School of Computer Science and Engineering, The University of New South Wales}
  \city{Sydney} 
  \state{NSW} 
  \postcode{2052}
    \country{Australia}
  }
\email{xianzhi.wang@unsw.edu.au}

\author{Wei Emma Zhang}
\affiliation{
  \institution{Department of Computing, Macquarie University}
  \city{Sydney} 
  \state{NSW} 
  \postcode{2109}
   \country{Australia}
}
\email{w.zhang@mq.edu.au}

\author{Anne H.H. Ngu}
\affiliation{
  \institution{Department of Computer Science, Texas State University}
  \city{San Marcos} 
  \state{TX} 
  \postcode{78666}
    \country{USA}
  }
\email{angu@txstate.edu}

\renewcommand{\shortauthors}{X. S. Fang et al.}
\renewcommand{\shorttitle}{Comparing Truth Discovery Methods without Using Ground Truth}

\begin{abstract}
Truth discovery has been widely studied in recent years as a fundamental means 
for resolving the conflicts in multi-source data. Although many truth discovery methods have been proposed based on different considerations and intuitions, investigations show that no 
single method consistently outperforms the others.
To select the right truth discovery method for a specific application scenario, it becomes essential to evaluate and compare the performance of different methods. 
A drawback of current research efforts is that they commonly assume the availability of certain ground truth for the evaluation of methods. However, the ground truth may be very limited or even out-of-reach in practice, rendering the evaluation biased by the small ground truth or even unfeasible. In this paper, we present \emph{CompTruthHyp}, a general approach for comparing the performance of truth discovery methods without using ground truth. In particular, 
our approach calculates the probability of observations in a dataset based on the output of different methods. The probability is then ranked to reflect the performance of these methods. We review and compare twelve existing truth discovery methods and consider both single-valued and multi-valued objects. Empirical studies on both real-world and synthetic datasets demonstrate the effectiveness of our approach for comparing truth discovery methods.
\end{abstract}



\maketitle

\section{Introduction}
The World Wide Web (WWW) has become a platform of paramount importance for storing, collecting, processing, querying, and managing the Big Data in recent years, with around $2.5$ quintillion bytes of data created every day through various channels such as blogs, social networks, discussion forums, and crowd-sourcing platforms. People from various domains, such as medical care, government, business, research, are relying on these data to fulfill their information needs. In these scenarios, information about the same objects can often be collected from a variety of sources. However, due to the varying quality of Web sources, the data about the same objects may conflict among the sources. To help users determine the veracity of multi-source data, a fundamental research topic, \emph{truth discovery}, has attracted broad attentions recently. 

So far, various truth discovery methods~\cite{li2015survey} have been proposed based on different considerations and intuitions. However, investigations~\cite{li2015survey,waguih2014truth,li2012truth} show that no methods could constantly outperform the others in all application scenarios. Moreover, Li et al.~\cite{li2012truth} demonstrate with experiments that even an improved method does not always beat its original version, such as \emph{Investment} and \emph{PooledInvestmen}~\cite{pasternack2010knowing}, \emph{Cosine, 2-Estimates} and \emph{3-Estimates}~\cite{galland2010corroborating}. Therefore, to help users select the most suitable method to fulfill their application needs, it becomes essential to evaluate and compare the performance of different methods. 

To evaluate the effectiveness of truth discovery methods,
current research usually measures their performance in terms of \emph{accuracy} (or \emph{error rate}), \emph{$F_1$-score}, \emph{recall}, \emph{precision}, \emph{specificity} for categorical data~\cite{waguih2014truth}, and \emph{Mean of Absolute Error} (MAE) and \emph{Root of Mean Square Error} (RMSE) for continuous data~\cite{li2015survey}. All these metrics are measured and compared based on the assumption that a reasonable amount of ground truth is available. However, the fact is, ground truth is often very limited or even out-of-reach (generally less than 10\% of the original dataset's size~\cite{waguih2014truth}). For example, the knowledge graph construction~\cite{Dong2014Vault} involves a huge number of objects, making it impossible to have the complete ground truth for performance validation. In addition, it requires enormous human efforts to acquire even a small set of ground truth. The lack of sufficient ground truth can, in many cases, statistically undermine the legitimacy of evaluating and comparing existing methods using the ground truth-based approach. For example, previous comparative studies~\cite{dong2009integrating,dong2010global,li2012truth,zhao2012probabilistic,zhao2012bayesian,li2014resolving,pochampally2014fusing,wang2015integrated,wang2016implications}, based on real-world datasets with sparse ground truth, could all bring bias to the performance measurement of the methods. Under this circumstance, it is hard to conclude which method performs better as we cannot trust the comparison results. 
This also makes it difficult to select the method with the best performance to be applied to specific application scenarios. Therefore, evaluating the performance of various truth discovery methods with missing or very limited ground truth can be a significant and challenging problem for the truth discovery applications~\cite{li2015survey}. We identify the key challenges of this issue as the following:
\begin{itemize}
\item The only way to obtain evidence for performance evaluation without ground truth is to extract features from the given dataset for truth discovery. However, the features of a dataset are sometimes complex, concerning source-to-source, source-to-object, object-to-value, and value-to-value relations. In addition, it is challenging to find a method to capture those relations without importing additional biases.
\item Current truth discovery methods commonly jointly determine value veracity and calculate source trustworthiness. Source trustworthiness and value confidence scores are the common intermediates of the existing methods, which are also the key elements for identifying the truth for each object. Therefore, we can consider identifying the relations among sources, objects, and values by leveraging those measurements to match the relations extracted from the given dataset. However, even if we are able to obtain the features of the given dataset, different truth discovery methods may calculate the source trustworthiness and value confidence scores using different metrics, which have various meanings and require non-trivial normalization tasks.  
\item Even if we are able to resolve the above two challenging issues, it is still tricky to find appropriate metrics for comparing those features, to fulfill the requirement of method comparison.
\end{itemize}

In this paper, we focus on truth discovery method comparison without using ground truth. In a nutshell, we make the following contributions in this paper:
\begin{itemize}
\item To our knowledge, we are the first to reveal the bias introduced by sparse ground truth in evaluating the truth discovery methods, by conducting experiments on synthetic datasets with different coverages of the leveraged ground truth. 
\item We 
analyze, implement, and compare twelve specific truth discovery methods, including \emph{majority voting, Sums, Average-Log, Investment, PooledInvestment}~\cite{pasternack2010knowing}, \emph{TruthFinder}~\cite{yin2008truth}, \emph{2-Estimates, 3-Estimates}~\cite{galland2010corroborating}, \emph{Accu}~\cite{dong2009integrating}, \emph{CRH}~\cite{li2014resolving}, \emph{SimpleLCA}, and \emph{GuessLCA}~\cite{pasternack2013latent}.
\item We propose a novel approach, called \emph{CompTruthHyp}, to \textbf{comp}are the performance of \textbf{truth} discovery methods without using ground truth, by considering the output of each method as a \textbf{hyp}othesis about the ground truth. CompTruthHyp takes both single-valued and multi-valued objects into consideration. It utilizes the output of all methods to quantify the probability of observation of the dataset and then determines the method with the largest probability to be the most accurate. 
\item We conduct extensive experiments on both synthetic and real-world datasets to demonstrate the effectiveness of our proposed approach. Our approach consistently achieves more accurate rankings of the twelve methods than traditional ground truth-based evaluation approach.
\end{itemize}

The rest of the paper is organized as follows. We review the related work in Section~\ref{sec:Related_Work}. Section~\ref{sec:Preliminaries} introduces some background knowledge about truth discovery and the observations that motivate our work. Section~\ref{sec:Approach} presents our approach. We report our experiments and results in Section~\ref{sec:Experiments}. Finally, Section~\ref{sec:Conclusion} provides some concluding remarks.

\section{Related Work}
\label{sec:Related_Work}
Generally, there are two categories of previous studies on performance evaluation and comparison of truth discovery methods. The first category includes the work on novel and advanced approaches for truth discovery in various scenarios. To validate the performance of their proposed approaches and to show how their approaches outperform the state-of-the-art truth discovery methods, those projects conduct comparative studies by running experiments on real-world datasets with manually collected ground truth. Truth discovery is first formulated by Yin et al.~\cite{yin2008truth}. To show the effectiveness of \emph{TruthFinder}, they conduct experiments on one real-world dataset, i.e., \emph{Book-Author} dataset, which contains $1,263$ objects. 
The manually collected ground truth 
only covers $7.91$\% of the objects. With truth discovery becomes more and more popular, considerable methods~\cite{pasternack2010knowing,dong2010global,dong2012less,zhao2012bayesian,zhao2012probabilistic,wang2015integrated,wang2016implications,Xiao2016Confidence,xiu2017value,Popat2017Where} have been proposed to fit various scenarios. We find that there is a common limitation of those works that they all conduct experiments on real-world datasets with limited ground truth. Besides the Book-Author dataset, the frequently-used datasets, including \emph{Flight}~\cite{li2012truth} (covers $8.33$\% of complete ground truth), \emph{Weather}~\cite{dong2010global} ($74.4$\%), \emph{Population}~\cite{pasternack2010knowing} ($0.702$\%), \emph{Movie}~\cite{zhao2012bayesian} ($0.663$\%) and \emph{Biography}~\cite{pasternack2010knowing} ($0.069$\%) all feature sparse or low-quality ground truth, which makes the experimental data evaluated on those datasets cannot be fully trusted. 

The second category of studies is presented in surveys~\cite{li2012truth,waguih2014truth,li2015survey} that aim at investigating and analyzing the strengths and limitations of the current state-of-the-art techniques. In particular, in 2012, Li et al.~\cite{li2012truth} studied the performance of sixteen data fusion methods in terms of precision and recall, on two real-world domains, namely \emph{Stock} and \emph{Flight}. 
Based on their experiments, the authors pointed out that 
the collected ground truth tends to trust data from certain sources, which sometimes puts wrong values or coarse-grained values in the ground truth. 
Moreover, we find that their constructed ground truth are relatively sparse, with the one for the stock domain covering only $200/1000=20$\% of the complete ground truth, and the one for the flight domain covering only $100/1200=8.33$\%. The most recent survey~\cite{li2015survey} provides a comprehensive overview of truth discovery methods and summarizes them from five different aspects, but they do not conduct any comparative experiments to show the diverse performance of the methods. Waguih et al.~\cite{waguih2014truth} point out that the sparse ground truth is not statistically significant to be legitimately leveraged for the accuracy evaluation and comparison of methods. To the best of our knowledge, they are the first to implement a dataset generator to generate synthetic datasets with the control over ground truth distribution, for the sake of comparing existing methods. Different from their work, our approach tries to evaluate the performance of various truth discovery methods without using ground truth, which is applicable to more general real-world scenarios.

\section{Preliminaries}
\label{sec:Preliminaries}
Current truth discovery methods take as input some conflicting triples (i.e., a given dataset) in the form of \{\emph{source, object, value}\}, where \emph{source} ($s \in S$) denotes the location where the data originates, \emph{object} ($o \in O$) is an attribute of an entity, and \emph{value} ($V_{s_o} \subset V$) depicts the potential value set of an object claimed by a source. For example, a triple, \{``www.imdb.com'', ``the director of Beauty and the Beast'', ``Bill Condon''\}, indicates that the website ``IMDb'' claims that the director of the movie ``Beauty and the Beast'' is ``Bill Condon''. If $o$ is a single-valued object, $|V_{s_o}|=1$. For example, ``the age of a person'' only has one single value; on the other hand, if $o$ is a multi-valued object, $|V_{s_o}|$ might be bigger than 1. For example, a person might have more than one child. Based on the triples, the methods infer a Boolean truth label (``true''/``false'') for each triple as the output. Formally, we name the facutal value of an object $o$ as the \emph{ground truth} of $o$, denoted by ${V_o}^*$, and the triple involves $o$ with the label ``true'' output by a truth discovery method $m$ as the \emph{identified truth} of $o$, denoted by ${V_o}^m$. After applying a group of truth discovery methods $M$ one by one on the triples, each method $m \in M$ outputs the \emph{identified truth} for each object $o \in O$. The closer ${V_o}^m$ is to ${V_o}^*$ for each object, the better the method $m$ performs. We denote the \emph{identified truth} of all objects in $O$ output by method $m$ as $V^m$ (${V_o}^m \subset V^m$), and the \emph{ground truth} of all objects in $O$, i.e., the \emph{complete ground truth} of the given dataset, as $V^*$ (${V_o}^* \subset V^*$).
In most cases, the ground truth provided with each frequently utilized real-world dataset, denoted by $V^i$, is only a subset of the complete ground truth ($V^i \subset V^*$). We define the \emph{coverage} of the ground truth as follows:

\begin{definition}{\textbf{Coverage of the Ground Truth}. } The percentage of objects covered by the ground truth over all the objects in the given dataset. The coverage of the complete ground truth is $100$\%.
$\square$
\end{definition}

\subsection{Ground Truth-Based Evaluation Approach}
\label{subsec:Observations}
Given the output of each truth discovery method, i.e., $V^m$, $m \in M$, and the ground truth ($V^i$), the traditional ground truth-based evaluation approach evaluates the effectiveness of each method in terms of \emph{precision, recall, F$_1$ score, accuracy/error Rate}, and \emph{specificity} for categorical data. For each metric of each method, the higher value is measured, the better the method performs. In particular, to derive those five metrics, the ground truth-based approach first produces a confusion matrix (as shown in Table~\ref{tab:confusion}) for each method. It cumulatively counts the numbers of true positives, false positives, true negatives, and false negatives for each object $o$ covered by $V^i$. Then, based on the matrix, it calculates the metrics as follows:

\begin{table}[]
\centering
\vspace{-2mm}
\caption{Confusion matrix of method $m$.}
\vspace{-2mm}
\label{tab:confusion}
\begin{tiny}
\begin{tabular}{|l|c|c|c|}
\hline
\multicolumn{2}{|l|}{\multirow{2}{*}{}}               & \multicolumn{2}{c|}{Ground Truth}         \\ \cline{3-4} 
\multicolumn{2}{|l|}{}                                & True                & False               \\ \hline
\multicolumn{1}{|c|}{\multirow{2}{*}{Method}} & True  & True Positive ($TP_m$)  & False Positive ($FP_m$) \\ \cline{2-4} 
\multicolumn{1}{|c|}{}                        & False & False Negative ($FN_m$) & True Negative ($TN_m$)  \\ \hline
\end{tabular}
\end{tiny}
\end{table}

\begin{itemize}
\item \emph{Precision} of method $m$ represents the probability of its positive outputs being correct, computed as $\frac{TP_m}{TP_m+FP_m}$.
\item \emph{Recall} of method $m$ indicates the probability of true values being identified as true, computed as $\frac{TP_m}{TP_m+FN_m}$. $1-recall$ is the so-called \emph{false negative rate}.
\item \emph{F$_1$ score} of method $m$ demonstrates the harmonious mean (i.e., a weighted average) of \emph{precsion} and \emph{recall}, computed as $\frac{2 \cdot precision \cdot recall}{precision + recall}$. 
\item \emph{Accuracy/Error Rate} of method $m$ is the probability of its outputs being correct, computed as $\frac{TP_m+TN_m}{TP_m+FP_m+TN_m+FN_m}$.
\item \emph{Specificity} of method $m$ presents the probability of false values being identified as false, computed as $\frac{TN_m}{FP_m+TN_m}$. $1-specificity$ is the so-called \emph{false positive rate}.
\end{itemize}

However, as $V^i$ is generally only a very small part of $V^*$, the distributions of true positives, false positives, true negatives, and false negatives, obtained in this small sample space cannot reflect the real distributions. Therefore, the derived metrics are not statistically significant to be legitimately leveraged for method accuracy evaluation and comparison. We will show the biases introduced by the limited ground truth in Section~\ref{subsec:Motivation}.

Additionally, most of the existing truth discovery methods assume that each object in the given dataset has only one true value~\cite{zhao2012bayesian}. When multi-valued objects (e.g., ``children of a person'') exist in the given dataset, they simply
concatenate and regard the values provided by the same source as a single joint value. 
Thus, under such assumption, by identifying a value of an object to be true, a truth discovery method is believed to implicitly claim that all the other values of the object are false. When a method incorrectly identifies a false value of an object to be true, it certainly asserts the true value as a false value. 
In this case, the false positives are equivalent to false negatives, and the recall and F$_1$ scores equal to the precision. 
However, when it comes to the case of multi-valued objects, the identified truth of a multi-valued object may overlap with the ground truth.
Simply labeling a value set as true or false according to whether it equals to the ground truth will degrade the accuracy of the performance evaluation of the method. For example, if the identified truth for ``Tom's children'' is \{``Anna, Tim''\}, and the ground truth is \{``Anna, Tim, Lucas''\}, the identified truth is partially true, rather than false. Therefore, we propose to treat each value in the identified value set individually. In this case, the false positives are no longer equivalent to false negatives. Neither the precision nor the recall of a method can reflect the performance of the method individually, we need to measure both the accuracy and the completeness of the methods' output. For example, given two methods $m_1$ and $m_2$, $m_1$ identifies \{``Anna, Tim''\} as ``Tom's children'', while $m_1$ identifies \{``Anna''\} is the only child of ``Tom''. The precision of both methods is $1$, as their identified values are all true values, indicating their performance are the same. However, the recall of $m_1$ is $\frac{2}{3}$ and that of $m_2$ is $\frac{1}{3}$, indicating the performance of $m_1$ is better than $m_2$.

In this paper, we will evaluate the performance of methods separately for \emph{single-valued scenario}s (i.e., scenarios where only single-valued objects exist) and \emph{multi-valued scenario}s (i.e., scenarios where multi-valued objects exist).

\subsection{Motivation}
\label{subsec:Motivation}
To investigate the bias introduced by the incomplete ground truth on method performance evaluation, we conducted experiments by evaluating twelve truth discovery methods (we will introduce these methods in Section~\ref{subsec:Twelve_Methods}), on synthetic datasets while tuning the \emph{coverage} of the ground truth.

\begin{figure}[!tb]
\centering
\subfloat[][\scriptsize{Synthetic80P Dataset}]{\includegraphics[width=1.8in]{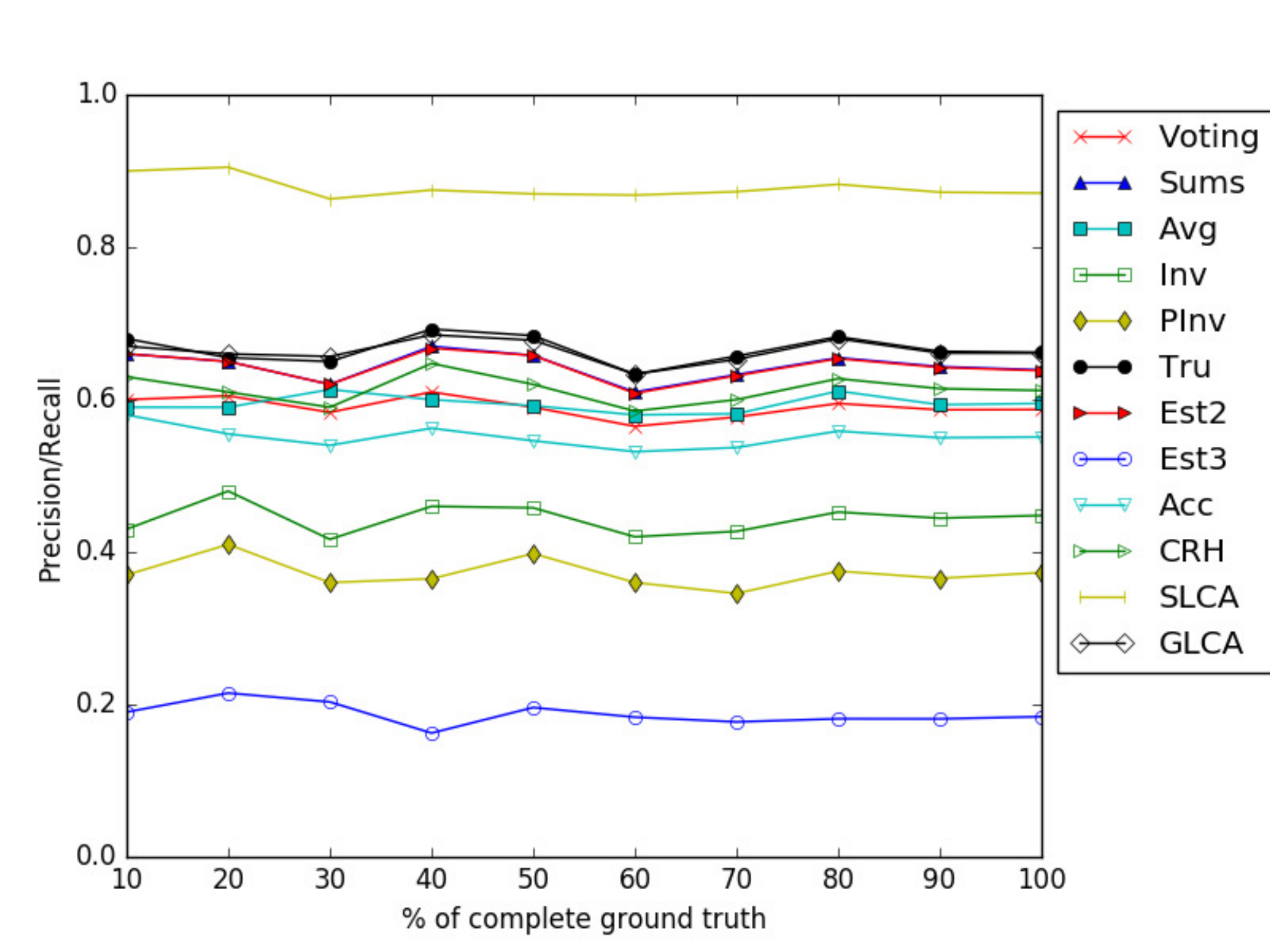}\label{fig:motivtation-a}}
~~
\subfloat[][\scriptsize{Synthetic80P Dataset}]{\includegraphics[width=1.8in]{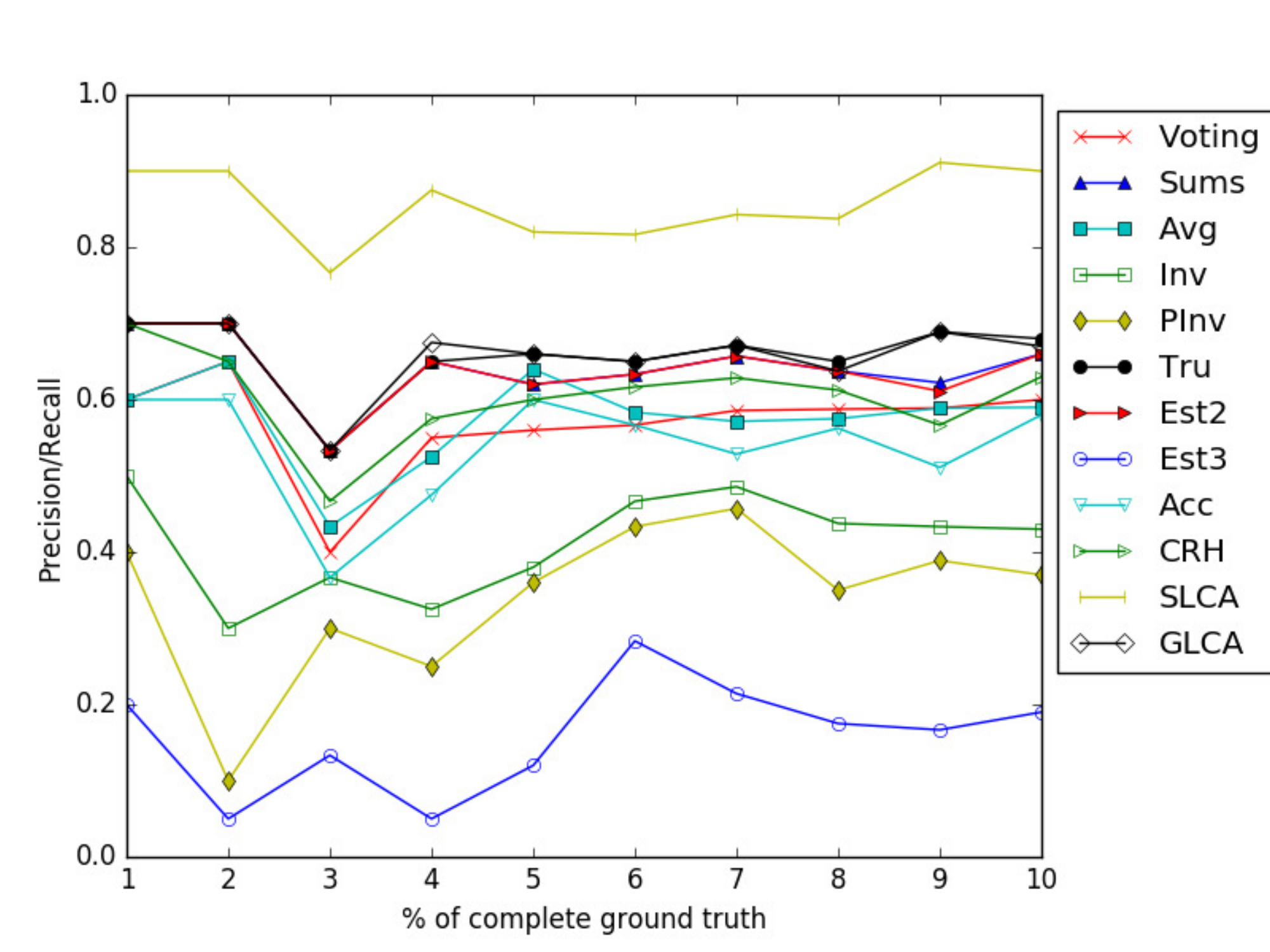}\label{fig:motivtation-b}}

\subfloat[][\scriptsize{SyntheticFP Dataset}]{\includegraphics[width=1.8in]{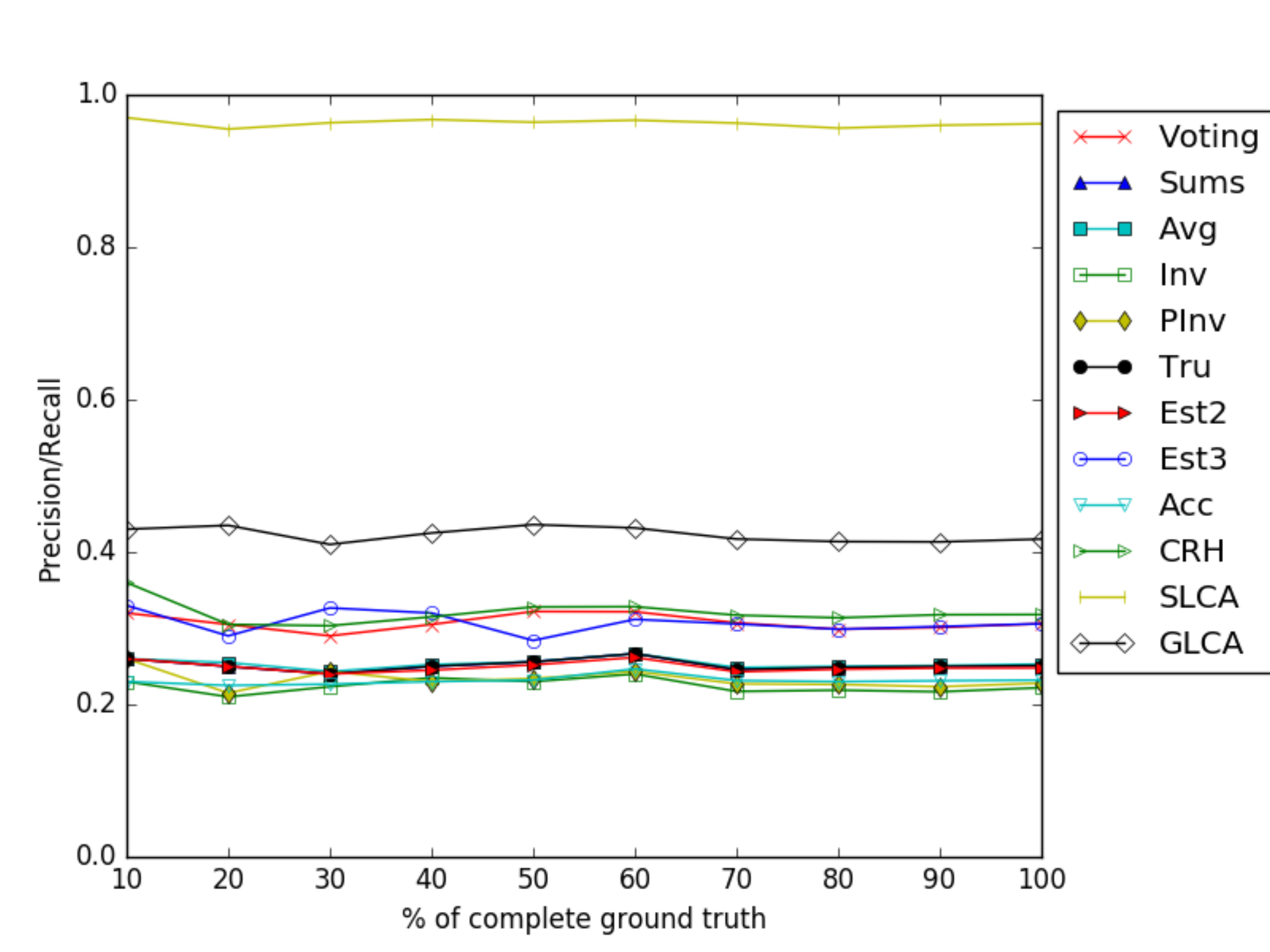}\label{fig:motivtation-c}}
~~
\subfloat[][\scriptsize{SyntheticFP Dataset}]{\includegraphics[width=1.8in]{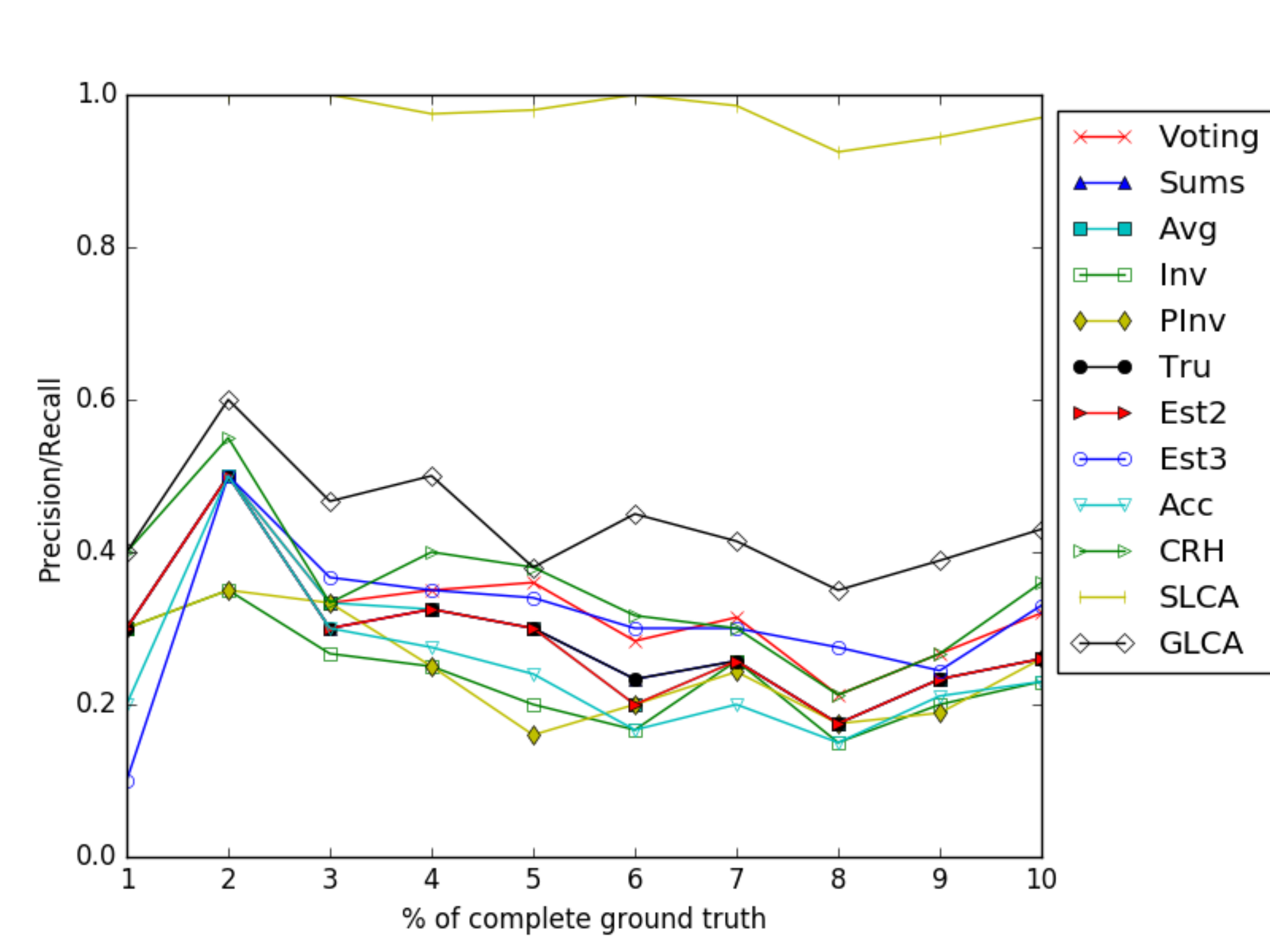}\label{fig:motivtation-d}}
\vspace{-2mm}
\caption{Precision/Recall of twelve truth discovery methods evaluated on different coverages of the leveraged ground truth}
\vspace{-2mm}
\label{fig:Statistics}
\end{figure}

The synthetic datasets with complete ground truth are generated by the dataset generator implemented by Waguih et al.~\cite{waguih2014truth}. 
This generator involves six parameters that are required to be configured to simulate a wide spectrum of truth discovery scenarios.  We will introduce the settings of those parameters in detail in Section~\ref{subsec:Experimental_Setup}. We tuned the ground truth distribution per source ($GT$) for all the seven possible distributions, including \emph{uniform, Random, Full-Pessimistic, Full-Optimistic, 80-Pessimistic, 80-Optimistic}, and \emph{Exponential}. Based on the above configurations, we obtained seven groups of datasets, each containing 10 datasets. The metrics, namely \emph{precision, recall, F$_1$ score, accuracy} and \emph{specificity} of each method were measured as the average of 10 executions over the 10 datasets included by the same dataset type. To calculate the metrics, for each dataset, we tuned the coverage of the ground truth from 10\% to 100\%, and also from 1\% to 10\%, by randomly picking up the specific quantity of objects from the complete ground truth. 
Due to the limited space, we only show the experimental results on two settings, namely \emph{80-Pessimistic} and \emph{Fully-Optimistic}, with the corresponding datasets 
depicted as \emph{Synthetic80P} and \emph{SyntheticFP}. 
The experimental results on all the other datasets show the same features. Note that all the objects in the synthetic datasets have only one true value, thus the resulting precision, recall, and F$_1$ score equal to each other. The accuracy and specificity show the same ranking features. Figure~\ref{fig:motivtation-a} and Figure~\ref{fig:motivtation-c}) show the precision and recall of all the twelve methods with the coverage of the leveraged ground truth tuned from 10\% to 100\%, while Figure~\ref{fig:motivtation-b} and Figure~\ref{fig:motivtation-d} show those of the methods with the coverage tuned from 1\% to 10\%.  The latter range forms the sparse ground truth, which is closer to the reality, where the coverage of the collected ground truth is always below 10\%, sometimes even below 1\%. 

Ideally, if the performance evaluation is not biased by the incomplete ground, there should be no intersecting lines in the figures, demonstrating that the ranking of the metrics of the methods is consistent with the results measured on complete ground truth. Even if two or more methods show the same performance, there should be overlapping lines rather than intersecting lines. However, for both types of datasets, we cannot get the completely correct ranking for each type of datasets until the coverage of the leveraged ground truth grew up to 60\%, which is generally impossible to obtain in reality. The results were even worse for the sparse ground truth. As shown in Figure~\ref{fig:motivtation-b} and Figure~\ref{fig:motivtation-d}, by tuning the coverage of the ground truth, the ranking of methods fluctuated all the time, and no correct result was returned. That means the performance evaluation is strongly biased by the sparse ground truth. In most cases, real-world datasets would not have strict mathematical distributions, such as source coverage distributions, ground truth distribution per source, and distinct value distribution per object might be random. Therefore, the ranking based on real-world datasets with sparse ground truth would be even less correct.

\section{Approach}
\label{sec:Approach}
The most straightforward approach for truth discovery is to conduct majority voting for categorical data or averaging for continuous data. The largest limitation of such approach is that it assumes all the sources are equally reliable, which does not hold in most real-world scenarios. Thus, the most important feature of the existing truth discovery methods is their ability to estimate source trustworthiness~\cite{li2015survey}. While identifying the truth, current methods also return $c_v$, the confidence score of each value $v$ (or the probability of $v$ being true), and $\tau_s$, the trustworthiness of each source $s$ (or the probability of source $s$ providing true information), as the intermediate variables. In particular, a higher $c_v$ indicates that value $v$ is more likely to be true, and a higher $\tau_s$ indicates that source $s$ is more reliable and the values claimed by this source are more likely to be true. Though the calculations of $c_v$ and $\tau_s$ differ from method to method, current methods generally apply the same principle for truth discovery: if a source claims true values frequently, it will receive a high trustworthiness; meanwhile, if a value is claimed by sources with high trustworthiness, it will be more likely to be identified as truth. To determine the truth, weighted aggregation of the multi-source data is performed based on the estimated source trustworthiness. Thus, value confidence score and source trustworthiness calculation are the key elements for truth discovery and can be leveraged to compare the performance of current truth discovery methods. In this section, we first review twelve existing truth discovery methods to be compared. Then we demonstrate our approach, \emph{CompTruthHyp}, which compares those methods without using ground truth in both single-valued scenarios and multi-valued scenarios. 

\subsection{Twelve Truth Discovery Methods}
\label{subsec:Twelve_Methods}
In this section, we describe each algorithm briefly.
\vspace{2mm}

\noindent \textbf{Majority voting. }By regarding all the sources as equally reliable, \emph{voting} does not estimate the trustworthiness of each source. Instead, it calculates $c_v$ as $\frac{|S_v|}{S_o}$, where $S_v$ is the set of sources which provide $v$ on object $o$, and $S_o$ is the set of sources which provide values on $o$. For each object, the value with the highest confidence score will be identified as the truth. 

\vspace{2mm}

\noindent \textbf{TruthFinder~\cite{yin2008truth}. }It applies Bayesian analysis to estimate source trustworthiness and identify the truth. It additionally takes value similarity into consideration. \emph{Truthfinder} terminates when the results from two successive iterations are less than a given threshold. The value with the highest confidence score is selected as the true value. 

\vspace{2mm}

\noindent \textbf{Accu~\cite{dong2009integrating}. }Dong et al. propose the first Bayesian truth detection model that incorporates copying detection techniques. Accu estimates trustworthiness of each source as the average confidence score of its provided values, while calculates value confidence scores by leveraging source trustworthiness using Bayes Rule. 

\vspace{2mm}

\noindent \textbf{Sums, Average-Log, Investment, PooledInvestment~\cite{pasternack2010knowing,pasternack2011making}. } \emph{Sums} employs \emph{authority-hub analysis}~\cite{kleinberg1999authoritative} and computes source trustworthiness as the average confidence score of its provided values. It has the disadvantage of overestimating the sources that make larger coverage of objects. \emph{Average-Log}, \emph{Investment}, and \emph{PooledInvestment} apply different methods to assess source trustworthiness. Specifically, Average-Log uses a non-linear function to assess sources. Investment assumes each source uniformly invests/distributes its trustworthiness to the values it provides, while sources collect credits back from the confidence of their claimed values. PooledInvestment follows a similar procedure with Investment, except it uses a linear function to estimate the confidence of values instead of a non-linear function.

\vspace{2mm}

\noindent \textbf{2-Estimates, 3-Estimates~\cite{galland2010corroborating}. }\emph{2-Estimates} incorporates the mutual exclusion (i.e., while claiming a value of an object, a source is voting against all the other potential values of this object). \emph{3-Estimates} augments 2-Estimates by additionally taking the hardness of fact into consideration.

\vspace{2mm}

\noindent \textbf{SimpleLCA, GuessLCA~\cite{pasternack2013latent}. }LCA (Latent Credibility Analysis) models source trustworthiness by a set of latent parameters. It enriches the meaning of source trustworthiness by tackling the difference between telling the truth and knowing the truth. In the \emph{SimpleLCA} model, source trustworthiness is considered as the probability of a source asserting the truth, while in the \emph{GuessLCA} model, source trustworthiness is regarded as the probability of a source both knowing and asserting the truth.

\vspace{2mm}

\noindent \textbf{CRH~\cite{li2014resolving}. }It is a framework that tackles heterogeneity of data. Source trustworthiness calculation is jointly conducted across all the data types together. Different types of distance functions can be incorporated into the framework to capture the features of different data types.

\subsection{CompTruthHyp}
\label{subsec:CompTruthHyp}
To compare the performance of the above methods without using the ground truth, our data model includes the following inputs: i) the input dataset for truth discovery; ii) the identified truth of each method ($m \in M$, $|M|=12$); iii) source trustworthiness and value confidence scores output by each method. The output of our data model is a ranking of the accuracy of the twelve methods. As we do not have any ground truth, we propose to obtain the ranking by comparing the methods' ability to infer the observation of the given dataset from their outputs. We denote by $\phi$ the observation of which source votes for which value in the dataset, $\phi(m)$ the output of a method $m$, and $P(\phi|\phi(m))$ the probability of $\phi$ conditioned on $\phi(m)$. A higher $P(\phi|\phi(m))$ indicates that the method $m$ has bigger ability to capture the features of the given dataset, thus its output is more reliable. 

Our computation requires several parameters, which can be derived from the inputs: ${\tau_s}(m)$, the probability that the claimed values of $s$ is true, given $V^m$. We will introduce the calculation of ${\tau_s}(m)$ in Section~\ref{subsec:Source_Trustworthiness_Normalization}; $P_o(v_t|V_o^m)$ (resp., $P_o(v_f|V_o^m)$), the probability that a source provides a particular true (resp., false) value for object $o$, given $V^m$. We will introduce the calculations of $P_o(v_t|V_o^m)$ and $P_o(v_f|V_o^m)$ in Section~\ref{subsec:Value_Distributions}. As analyzed in Section~\ref{subsec:Observations}, we compute the required parameters by applying different algorithms for single-valued and multi-valued scenarios. 

Given $V^m$, if $v \in V_o^m$, $v$ is a true value; if $v \in V_o-V_o^m$, $v$ is a false value. Formally, if a source $s$ covers an object $o$, we have the probability of the observation of $s$ providing a particular value $v$ ($v \in V_o$), conditioned on $\phi(m)$, as:

\begin{equation}
\label{equa:1}
P(\phi_{s_v}|\phi(m))= \left\{\begin{matrix}
 {\tau_s}(m) P_o(v_t|V^m);  & if v \in V_o^m \\ 
 (1-{\tau_s}(m)) P_o(v_f|V^m);& if v \in V_o-V_o^m 
\end{matrix}\right.
\end{equation}

In our observation, we are interested in two sets of values: given $V^m$, ${{V_s}^t}(m)$, denoting the set of true values provided by $s$; ${{V_s}^f}(m)$, denoting the set of false values provided by $s$. ${{V_s}^t}(m) \cup {{V_s}^f}(m) = V_s$, $V_s$ is the set of all values provided by $s$. Since we assume each source provides each value independently, we have the probability of the observation of source $s$ with its claimed values, i.e., $\phi_s$, conditioned on $\phi(m)$, as:

\begin{tiny}
 \begin{equation}
\label{equa:2}
P(\phi_s|\phi(m)) = (\prod_{v \in {{V_s}^t}(m)}{\tau_s}(m) P_o(v_t|V^m)  \prod_{v \in {{V_s}^f}(m)}(1-{\tau_s}(m)) P_o(v_f|V^m))
\end{equation}
\end{tiny}

By assuming sources are independent on each other, the conditional probability of observing the given dataset $\phi$ is:

\begin{tiny}
 \begin{equation}
\label{equa:3}
P(\phi|\phi(m)) = \prod_{s \in S}{(\prod_{v \in {{V_s}^t}(m)}{\tau_s}(m) P_o(v_t|V^m)  \prod_{v \in {{V_s}^f}(m)}(1-{\tau_s}(m)) P_o(v_f|V^m))}
\end{equation}
\end{tiny}

To simplify the computation, we define the \emph{confidence} of method $m$, denote by $\mathcal C_m$, as 
\begin{tiny}
 \begin{equation}
\label{equa:4}
\mathcal C_m= \sum_{s \in S}{(\sum_{v \in {{V_s}^t}(m)}\ln {\tau_s}(m) P_o(v_t|V^m) + \sum_{v \in {{V_s}^f}(m)}\ln (1-{\tau_s}(m)) P_o(v_f|V^m))}
\end{equation}
\end{tiny}

\subsubsection{Source Trustworthiness Normalization}
\label{subsec:Source_Trustworthiness_Normalization}
The accuracy of truth discovery methods significantly depends on their ability of source trustworthiness estimation. Although all methods calculate source trustworthiness as the weighted aggregation of value confidence scores, they adopt different models and equations. Therefore, the calculated source trustworthiness of each method has different meaning and is incomparable with one another. To normalize source trustworthiness output by twelve methods, our approach, \emph{CompTruthHyp}, regards the trustworthiness of a source as the probability of its claimed values being true (i.e., precision). We can derive a confusion matrix similar to Table~\ref{tab:confusion} for each source based on the identified truth of each method. Then, we calculate the precision of each source output by each method (${\tau_s}(m)$) as follows:
\begin{equation}
\label{equa:5}
{\tau_s}(m) = \frac{{TP_s}^m}{{TP_s}^m+{FP_s}^m}
\end{equation}
where ${TP_s}^m$ (resp., ${FP_s}^m$) is the number of true positives (resp., false positives) of the values claimed by source $s$, given $V^m$.

In single-valued scenarios, each source provides one value for any object of interest. Given $V^m$, all the values in $V_o-{V_o}^m$ are regarded as false ($|V_o-{V_o}^m|=|V_o|-1$). We calculate ${\tau_s}(m)$ for each source by performing Algorithm~\ref{alg:1}. In particular, for each method $m \in M$ (Line $1$), for each $s \in S$ (Line $2$), for each $o \in O_S$ (Line $4$, where $O_s$ is the objects covered by $s$), if $V_{s_o}$ is true (Line $5$), ${TP_s}^m$ increases by one (Line $6$),  
otherwise, ${FP_s}^m$ increases by one (Line $7$, $8$). For each source $s$, ${\tau_s}(m)$ is calculated by applying Equation~\ref{equa:5} (Line $9$).

\begin{algorithm}[!th]
\begin{tiny}
\KwIn{Given dataset \{$s, o, V_{s_o}$\} and $V^m$ for each $m \in M$.}
\KwOut{${\tau_s}(m)$ for each $s \in S$, $m \in M$.}

\ForEach{$m\in M$}{
  \ForEach{$s\in S$}{
           {${TP_s}^m = 0$;${FP_s}^m = 0$;}\\
    \ForEach{$o \in O_s$}{
        \If {$V_{s_o}={V_o}^m$}{
        {${TP_s}^m++$;}\\
        }
        \Else{
           {${FP_s}^m++$;}\\
        }        
    }
    {Calculate ${\tau_s}(m)$  by applying Equation~\ref{equa:5};}\\
 }
}
\Return ${\tau_s}(m)$ for each $s \in S$, $m \in M$;\\
\caption{The algorithm of source trustworthiness normalization for single-valued scenarios.}
\label{alg:1}
\end{tiny}
\end{algorithm}

In multi-valued scenarios, each source may provide more than one value for a multi-valued object. Instead of regarding each value set provided by a source on the same object as a joint single value, we treat each value in the value set individually. Therefore, $|{V_o}^m|$ and $|V_{s_o}|$ may be bigger than $1$. We calculate ${\tau_s}(m)$ for each source by performing Algorithm~\ref{alg:2}.

\begin{algorithm}[!th]
\begin{tiny}
\KwIn{Given dataset \{$s, o, V_{s_o}$\} and $V^m$ for each $m \in M$.}
\KwOut{${\tau_s}(m)$ for each $s \in S$, $m \in M$.}

\ForEach{$m\in M$}{
  \ForEach{$s\in S$}{
           {${TP_s}^m = 0$;${FP_s}^m = 0$;}\\                                
    \ForEach{$o \in O$}{
       \ForEach{$v in V_{s_o}$} {       
             \If {$v \in {V_o}^m$}{                    
                        {${TP_s}^m++$;}\\
                        }
             \Else{
                    {${FP_s}^m++$;}\\
        }
        }
    }
    {Calculate ${\tau_s}(m)$  by applying Equation~\ref{equa:5};}\\
 }
}
\Return ${\tau_s}(m)$ for each $s \in S$, $m \in M$;\\
\caption{The algorithm of source trustworthiness normalization for multi-valued scenarios.}
\label{alg:2}
\end{tiny}
\end{algorithm}

\subsubsection{Value True-False Distributions}
\label{subsec:Value_Distributions}
In this section, we analyze the true-false distribution of values for each object in a given dataset. Because each object has one single value in single-valued scenarios, we have $P_o(v_t|V^m)$ fixed to $1$. As false values for an object can be random, $P_o(v_f|V^m)$ is different for false values, and the false value distribution of object is also different. Given a set of false values of $o$ ($V_o-V_o^m$), we need to analyze their distribution and calculate the probability ($P_o(v_f|V^m)$) for sources to pick a particular value from the distribution. In particular, we calculate this probability for each false value of each object using Algorithm~\ref{alg:3}. We define the \emph{untrustworthiness} of a source as the probability that its claimed values being false, i.e., $(1-{\tau_s}(m))$. For each false value of each object, we calculate $P_o(v_f|V^m)$ by:
\begin{equation}
\label{equa:6}
P_o(v_f|V^m)= \frac{\sum_{s \in S_{v_f}}{(1-\tau_s(m))}}{\sum_{{v_f}' \in V_o-V_o^m}{\sum_{s' \in S_{{v_f}'}}{(1-\tau_{s'}(m))}}}
\end{equation}
where $S_{v_f}$ is the set of sources provide $v_f$ on $o$.

\begin{algorithm}[!th]
\begin{tiny}
\KwIn{Given dataset \{$s, o, V_{s_o}$\} and $V^m$ for each $m \in M$.}
\KwOut{$P_o(v_f|V^m)$ for each $v_f \in V_o-V_o^m$, $o \in O$, $m \in M$.}

\ForEach{$m\in M$}{
    \ForEach{$o \in O$}{
           \ForEach{$v_f  \in V_o-V_o^m$}{
                \ForEach{$s \in S_{v_f}$}{
                     {$P_o(v_f|V^m)+=(1-{\tau_s}(m))$;}\\
                } 
             {$P_o(v_f|V^m)$ of each $v_f$ is normalized to satisfy $\sum_{v_f \in V_o-V_o^m}P_o(v_f|V^m)=1$;}\\                      
    }
    }
}
\Return{$P_o(v_f|V^m)$ for each $v_f \in V_o-V_o^m$, $o \in O$, $m \in M$;}\\
\caption{The algorithm of $P_o(v_f|V^m)$ calculation for single-valued scenarios.}
\label{alg:3}
\end{tiny}
\end{algorithm}

In multi-valued scenarios, the values in a source's claimed value set are not totally independent. Intuitively, the values occurring in the same claimed value set are believed to impact each other. The co-occurrence of values in the same claimed value set indicates a potentially similar probability of being selected.

\begin{figure}[!tb]
\centering
\subfloat[][\scriptsize{True value graph}]{\includegraphics[width=1.5in]{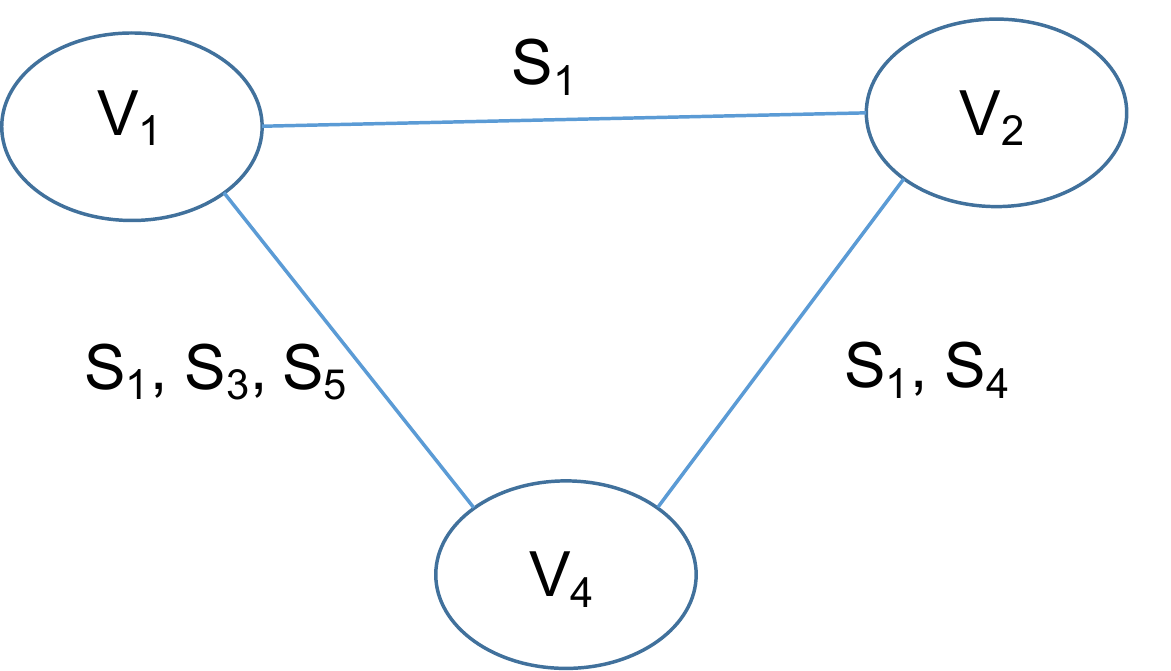}\label{fig:true_value_graph}}
~~
\subfloat[][\scriptsize{False value graph}]{\includegraphics[width=1.5in]{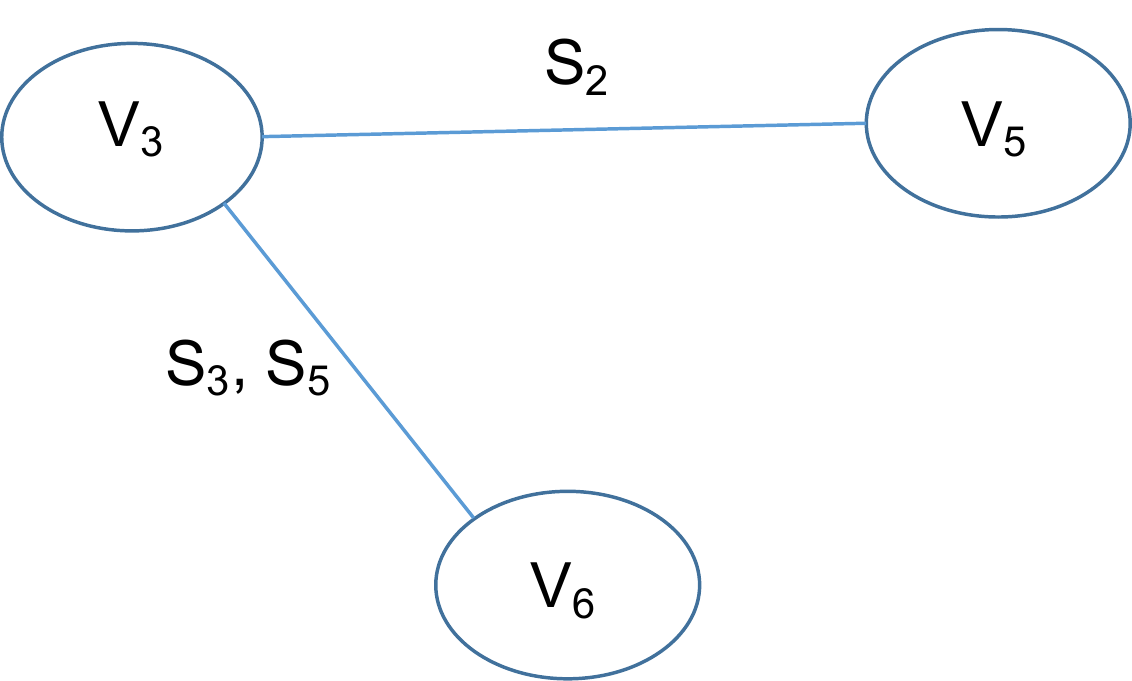}\label{fig:false_value_graph}}
\vspace{-2mm}
\caption{An example of value co-occurrences for a multi-valued object.}
\vspace{-2mm}
\label{fig:value_graph}
\end{figure}

We define the weighted association among the distinctive values on the same object to represent their influence to each other, based on which to compute the probability of each value being selected. In particular, given $V_o^m$, we represent the bipartite mapping between true (resp., false) values on each multi-valued object and sources that claim the true (resp., false) values into a true (resp., false) value graph. In each true (reps., false) value graph, the identified true values (resp., false values) in $V_o^m$ (resp., $V_o-V_o^m$) are the vertices, and sources that claim those values are the weights of edges which connect with the values. For example, the value co-occurrences for a multi-valued object are shown in Figure~\ref{fig:value_graph}. $V_o$ = \{$v_1$, $v_2$, $v_3$, $v_4$, $v_5$, $v_6$\}, $V_o^m$ = \{$v_1$, $v_2$, $v_4$\}. 
True values $v_2$ and $v_4$ are claimed by both $s_1$ and $s_4$, while false values $v_3$ and $v_5$ are claimed by $s_2$.

The detailed procedure of $P_o(v_t|V^m)$ and $P_o(v_f|V^m)$ calculation is shown in Algorithm~\ref{alg:4}. For each true (resp., false) value graph, we further generate a corresponding square adjacent ``true'' (resp., ``false'') matrix, which should be irreducible, aperiodic, and stochastic to be guaranteed to converge to a stationary state. In particular, we first initialize each element in the matrix as the sum of the trustworthiness (resp., untrustworthiness) of all sources that claim the co-occurrence of the corresponding pair of true (resp.,  false) values (Line $8$ and Line $17$). To guarantee the three features of the matrix, we add a ``\emph{smoothing link}'' by assigning a small weight to every pair of values (Line $9$ and Line $18$), where $\beta$ is the smoothing factor. For our experiments, we set $\beta=0.1$ (empirical studies such as the work done by Gleich et al.~\cite{Gleich2010surfer} demonstrate more accurate estimation).  We then normalize the elements to ensure that every column in the matrix sums to $1$ (Line $10$ and Line $19$). This normalization allows us to interpret the elements as the transition probabilities for the random walk computation. Finally, we adopt the \emph{Fixed Point Computation Model} (FPC)~\cite{brin1998pagerank} on each ``true'' (resp., ``false'') matrix to calculate $P_o(v_t|V^m)$ (reps., $P_o(v_f|V^m)$) for each true (resp., false) value of each object $o \in O$ (Line $11$ and Line $20$). 

\begin{algorithm}[!th]
\begin{tiny}
\KwIn{Given dataset \{$s, o, V_{s_o}$\} and $V^m$ for each $m \in M$.}
\KwOut{$P_o(v_t|V^m)$ for each $v_t \in V_o^m$, $P_o(v_f|V^m)$ for each $v_f \in V_o-V_o^m$, $o \in O$, $m \in M$.}
{$\beta = 0.1$;}\\
\ForEach{$m\in M$}{
\tcp{``true'' matrix generation}
    \ForEach{$o \in O$}{
           \ForEach{$v_{t_1}  \in V_o^m$}{
                \ForEach{$v_{t_2}  \in V_o^m$}{
                \If {$v_{t_1}\neq v_{t_2}$} {
                \ForEach{$s \in S_{v_{t_1}} \cap S_{v_{t_2}}$}{
                { $TrueMatrix[v_{t_1}][v_{t_2}]+={\tau_s}(m)$;}\\
                 }
                  {$TrueMatrix[v_{t_1}][v_{t_2}] = \beta + (1-\beta)*TrueMatrix[v_{t_1}][v_{t_2}]$;}   \\                                                          
    } 
    }                                     
    }
    {Normalize TrueMatrix;}\\
    {Apply FPC random walk computation to obtain $P_o(v_t|V^m)$ for each $v_t \in V_o^m$;}\\
    }
\tcp{``false'' matrix generation}
    \ForEach{$o \in O$}{
           \ForEach{$v_{f_1} \in  V_o-V_o^m$}{
                \ForEach{$v_{f_2} \in  V_o-V_o^m$}{
                \If {$v_{f_1}\neq v_{f_2}$} {
                \ForEach{$s \in S_{v_{f_1}} \cap S_{v_{f_2}}$}{
                 {$FalseMatrix[v_{f_1}][v_{f_2}]+=1-{\tau_s}(m)$;}\\
                 }
                  {$FalseMatrix[v_{f_1}][v_{f_2}] = \beta + (1-\beta)*FalseMatrix[v_{f_1}][v_{f_2}]$;}   \\                                                          
    } 
    }                                     
    }
    {Normalize FalseMatrix;}\\
    {Apply FPC random walk computation to obtain $P_o(v_f|V^m)$ for each $v_f \in V_o-V_o^m$;}\\
    }
}
\Return $P_o(v_t|V^m)$ for each $v_t \in V_o^m$, $P_o(v_f|V^m)$ for each $v_f \in V_o-V_o^m$, $o \in O$, $m \in M$;\\
\caption{The algorithm of $P_o(v_t|V^m)$ and $P_o(v_f|V^m)$ calculation for multi-valued scenarios.}
\label{alg:4}
\end{tiny}
\end{algorithm}
\begin{table*}[]
\vspace{-2mm}
\centering
\vspace{-2mm}
\caption{Experimental results for six types of representative synthetic datasets (single-valued scenarios).}
\label{tab:Synthetic}
\begin{tiny}
\begin{tabular}{|p{5mm}|p{5mm}|l|p{4mm}|p{4mm}|l|p{4mm}|p{4mm}|p{4mm}|p{4mm}|p{4mm}|p{4mm}|p{4mm}|p{4mm}|p{4mm}|p{4mm}|p{4mm}|p{4mm}|p{4mm}|p{4mm}|p{5mm}|p{5mm}|}
\hline
  \textbf{Dataset}    & \textbf{Method} &\textbf{ P(1\%)}         & \textbf{P(2\%)}         & \textbf{P(3\%) }        &\textbf{ P(3\%)}        & \textbf{P(5\%)        } & \textbf{P(6\%)}         & \textbf{P(7\%)}         & \textbf{P(8\%) }        & \textbf{P(9\%)}         & \textbf{P(10\%)  }      & \textbf{P(20\%) }       & \textbf{P(30\%)}        & \textbf{P(40\%)}        & \textbf{P(50\%)}        & \textbf{P(60\%)}                      & \textbf{P(70\%) }       & \textbf{P(80\%) }       & \textbf{P(90\%) }       & \textbf{P(100\%)}       & \textbf{$\mathcal C_m$}            \\\hline
\multicolumn{1}{|c|}{}                       & Voting & 0.600          & 0.650          & 0.400          & 0.550          & 0.560          & 0.567          & 0.586          & 0.588          & 0.589          & 0.600          & 0.605          & 0.583          & 0.610          & 0.590          & 0.565                        & 0.577          & 0.595          & 0.587          & 0.587          & -16604          \\
\multicolumn{1}{|c|}{}                       & Sums   & 0.700          & 0.700          & 0.533          & 0.650          & 0.620          & 0.633          & 0.657          & 0.638          & 0.622          & 0.660          & 0.650          & 0.620          & 0.670          & 0.658          & 0.610                        & 0.633          & 0.655          & 0.643          & 0.639          & -16514          \\
\multicolumn{1}{|c|}{}                       & Avg    & 0.600          & 0.650          & 0.433          & 0.525          & 0.640          & 0.583          & 0.571          & 0.575          & 0.589          & 0.590          & 0.590          & 0.613          & 0.600          & 0.592          & 0.580                        & 0.581          & 0.611          & 0.593          & 0.595          & -16603          \\
\multicolumn{1}{|c|}{}                       & Inv    & 0.500          & 0.300          & 0.367          & 0.325          & 0.380          & 0.467          & 0.486          & 0.438          & 0.433          & 0.430          & 0.480          & 0.417          & 0.460          & 0.458          & 0.420                        & 0.427          & 0.453          & 0.444          & 0.448          & -17319          \\
\multicolumn{1}{|c|}{}                       & PInv   & 0.400          & 0.100          & 0.300          & 0.250          & 0.360          & 0.433          & 0.457          & 0.350          & 0.389          & 0.370          & 0.410          & 0.360          & 0.365          & 0.398          & 0.360                        & 0.346          & 0.375          & 0.366          & 0.373          & -17843          \\
\multicolumn{1}{|c|}{}                       & Tru    & 0.700          & 0.700          & 0.533          & 0.650          & 0.660          & 0.650          & 0.671          & 0.650          & 0.689          & 0.680          & 0.655          & 0.650          & 0.693          & 0.684          & 0.633                        & 0.657          & 0.683          & 0.663          & 0.662          & -16489          \\
\multicolumn{1}{|c|}{}                       & Est2   & 0.700          & 0.700          & 0.533          & 0.650          & 0.620          & 0.633          & 0.657          & 0.638          & 0.611          & 0.660          & 0.650          & 0.620          & 0.668          & 0.658          & 0.608                        & 0.631          & 0.654          & 0.642          & 0.638          & -16514          \\
\multicolumn{1}{|c|}{}                       & Est3   & 0.200          & 0.05           & 0.133          & 0.050          & 0.120          & 0.283          & 0.214          & 0.175          & 0.167          & 0.190          & 0.215          & 0.203          & 0.163          & 0.196          & 0.183                        & 0.177          & 0.181          & 0.181          & 0.184          & -18629          \\
\multicolumn{1}{|c|}{}                       & Accu   & 0.600          & 0.600          & 0.367          & 0.475          & 0.600          & 0.567          & 0.529          & 0.563          & 0.511          & 0.580          & 0.555          & 0.540          & 0.563          & 0.546          & 0.532                        & 0.537          & 0.559          & 0.550          & 0.551          & -16640          \\
\multicolumn{1}{|c|}{}                       & CRH    & 0.700          & 0.650          & 0.467          & 0.575          & 0.600          & 0.617          & 0.629          & 0.613          & 0.567          & 0.630          & 0.610          & 0.590          & 0.648          & 0.620          & 0.585                        & 0.600          & 0.628          & 0.614          & 0.612          & -16558          \\
\multicolumn{1}{|c|}{}                       & SLCA   & \textbf{0.900} & \textbf{0.900} & \textbf{0.767} & \textbf{0.875} & \textbf{0.820} & \textbf{0.817} & \textbf{0.843} & \textbf{0.838} & \textbf{0.911} & \textbf{0.900} & \textbf{0.905} & \textbf{0.863} & \textbf{0.875} & \textbf{0.870} & \textbf{0.868}               & \textbf{0.873} & \textbf{0.883} & \textbf{0.872} & \textbf{0.871} & \textbf{-15933} \\
\multicolumn{1}{|c|}{}                       & GLCA   & 0.700          & 0.700          & 0.533          & 0.675          & 0.660          & 0.650          & 0.671          & 0.638          & 0.689          & 0.670          & 0.660          & 0.657          & 0.685          & 0.678          & 0.633                        & 0.653          & 0.680          & 0.661          & 0.661          & -16480          \\\cline{2-22}
\multicolumn{1}{|c|}{}                       & \textbf{Dist.}  & 5.916          & 4.359          & 3.873          & 3.000          & 4.123          & 1.732          & 2.000          & 2.646          & 3.162          & 1.732          & 2.236          & 2.236          & 1.414          & 1.000          & 1.000                        & 0.000          & 0.000          & 0.000          & 0.000          & 1.414           \\
\multicolumn{1}{|c|}{\multirow{-14}{*}{\textbf{80P}}} & \textbf{Cos.}   & 0.975          & 0.987          & 0.989          & 0.993          & 0.987          & 0.998          & 0.997          & 0.995          & 0.992          & 0.998          & 0.996          & 0.996          & 0.998          & 0.999          & 0.999                        & 1.000          & 1.000          & 1.000          & 1.000          & 0.998           \\\hline\hline
\multicolumn{1}{|c|}{}                       & Voting & \textbf{1.000} & \textbf{0.950} & \textbf{1.000} & \textbf{1.000} & \textbf{1.000} & 0.983          & \textbf{1.000} & \textbf{1.000} & 0.989          & 0.990          & 0.985          & 0.990          & 0.983          & 0.992          & 0.992                        & 0.989          & 0.991          & 0.991          & 0.992          & -10574           \\
\multicolumn{1}{|c|}{}                       & Sums   & \textbf{1.000} & \textbf{0.950} & \textbf{1.000} & \textbf{1.000} & \textbf{1.000} & 0.983          & \textbf{1.000} & \textbf{1.000} & \textbf{1.000} & \textbf{1.000} & 0.985          & 0.990          & 0.985          & 0.992          & \textbf{0.993}               & 0.991          & 0.994          & 0.993          & 0.994          & -10567          \\
\multicolumn{1}{|c|}{}                       & Avg    & \textbf{1.000} & \textbf{0.950} & \textbf{1.000} & \textbf{1.000} & \textbf{1.000} & 0.983          & \textbf{1.000} & \textbf{1.000} & 0.989          & 0.990          & 0.985          & 0.990          & 0.983          & 0.992          & 0.992                        & 0.989          & 0.991          & 0.991          & 0.992          &-10574         \\
\multicolumn{1}{|c|}{}                       & Inv    & 0.900          & 0.800          & \textbf{1.000} & 0.875          & 0.800          & 0.817          & 0.871          & 0.900          & 0.889          & 0.900          & 0.815          & 0.833          & 0.813          & 0.834          & 0.837                        & 0.834          & 0.859          & 0.844          & 0.847          & -11944         \\
\multicolumn{1}{|c|}{}                       & PInv   & \textbf{1.000} & 0.850          & \textbf{1.000} & 0.925          & 0.940          & 0.850          & 0.914          & 0.950          & 0.944          & 0.960          & 0.880          & 0.897          & 0.868          & 0.888          & 0.892                        & 0.890          & 0.905          & 0.893          & 0.898          & -11537         \\
\multicolumn{1}{|c|}{}                       & Tru    & \textbf{1.000} & \textbf{0.950} & \textbf{1.000} & \textbf{1.000} & \textbf{1.000} & \textbf{1.000} & \textbf{1.000} & \textbf{1.000} & 0.989          & 0.990          & 0.990          & \textbf{0.997} & 0.990          & 0.994          & 0.992                        & 0.993          & 0.994          & 0.994          & 0.995          & -10554          \\
\multicolumn{1}{|c|}{}                       & Est2   & \textbf{1.000} & \textbf{0.950} & \textbf{1.000} & \textbf{1.000} & \textbf{1.000} & \textbf{1.000} & \textbf{1.000} & \textbf{1.000} & \textbf{1.000} & \textbf{1.000} & 0.990          & \textbf{0.997} & 0.990          & 0.994          & \textbf{0.993}               & \textbf{0.994} & \textbf{0.995} & \textbf{0.996} & \textbf{0.996} & -10554           \\
\multicolumn{1}{|c|}{}                       & Est3   & 0.200          & 0.150          & 0.200          & 0.125          & 0.200          & 0.150          & 0.257          & 0.138          & 0.133          & 0.250          & 0.155          & 0.210          & 0.195          & 0.192          & 0.212                        & 0.184          & 0.206          & 0.193          & 0.196          & -12107          \\
\multicolumn{1}{|c|}{}                       & Accu   & \textbf{1.000} & \textbf{0.950} & \textbf{1.000} & \textbf{1.000} & \textbf{1.000} & \textbf{1.000} & \textbf{1.000} & \textbf{1.000} & \textbf{1.000} & \textbf{1.000} & 0.990          & \textbf{0.997} & 0.990          & 0.994          & \textbf{0.993}               & \textbf{0.994} & \textbf{0.995} & \textbf{0.996} & \textbf{0.996} & -10554          \\
\multicolumn{1}{|c|}{}                       & CRH    & \textbf{1.000} & \textbf{0.950} & \textbf{1.000} & \textbf{1.000} & \textbf{1.000} & 0.983          & \textbf{1.000} & \textbf{1.000} & 0.989          & \textbf{1.000} & 0.985          & 0.993          & 0.983          & 0.986          & 0.990                        & 0.989          & 0.991          & 0.991          & 0.992          & -10577         \\
\multicolumn{1}{|c|}{}                       & SLCA   & \textbf{1.000} & \textbf{0.950} & \textbf{1.000} & \textbf{1.000} & \textbf{1.000} & \textbf{1.000} & \textbf{1.000} & \textbf{1.000} & 0.989          & 0.990          & \textbf{0.995} & \textbf{0.997} & \textbf{0.993} & \textbf{0.996} & \textbf{0.993}               & \textbf{0.994} & \textbf{0.995} & \textbf{0.996} & \textbf{0.996} & \textbf{-10552}  \\
\multicolumn{1}{|c|}{}                       & GLCA   & \textbf{1.000} & \textbf{0.950} & \textbf{1.000} & \textbf{1.000} & \textbf{1.000} & 0.967          & \textbf{1.000} & \textbf{1.000} & 0.978          & 0.990          & 0.980          & 0.990          & 0.980          & 0.990          & 0.990                        & 0.987          & 0.990          & 0.990          & 0.991          & -10578         \\\cline{2-22}
\multicolumn{1}{|c|}{}                       & \textbf{Dist.}  & 15.427         & 12.530         & 12.530         & 12.530         & 12.530         & 3.610          & 12.530         & 12.530         & 5.477          & 8.426          & 3.162          & 4.796          & 2.646          & 4.359          & 4.123 & 1.000          & 0.000          & 1.000          & 0.000          & 3.317          \\
\multicolumn{1}{|c|}{\multirow{-14}{*}{\textbf{80O}}} & \textbf{Cos.}   &0.778                &   0.859             &    0.859            &   0.859             &               0.859 &    0.989           &  0.859              &    0.859           &   0.975             &    0.938            &   0.992             &    0.981          &   0.994             &       0.984         &     0.985                        &     0.999          &    1.000            &    0.999            &      1.000          &  0.991           \\\hline\hline
                                             & Voting & 0.300          & 0.500          & 0.333          & 0.350          & 0.360          & 0.283          & 0.314          & 0.213          & 0.267          & 0.320          & 0.305          & 0.290          & 0.305          & 0.322          & 0.322                        & 0.307          & 0.299          & 0.301          & 0.306          & -19758                \\
                                             & Sums   & 0.300          & 0.500          & 0.300          & 0.325          & 0.300          & 0.233          & 0.257          & 0.175          & 0.233          & 0.260          & 0.250          & 0.240          & 0.250          & 0.256          & 0.267                        & 0.246          & 0.249          & 0.250          & 0.251          &   -19647              \\
                                             & Avg    & 0.300          & 0.500          & 0.333          & 0.325          & 0.300          & 0.200          & 0.257          & 0.175          & 0.233          & 0.260          & 0.255          & 0.243          & 0.253          & 0.256          & 0.267                        & 0.249          & 0.250          & 0.251          & 0.253          &   -19660            \\
                                             & Inv    & 0.300          & 0.350          & 0.267          & 0.250          & 0.200          & 0.167          & 0.257          & 0.150          & 0.200          & 0.230          & 0.210          & 0.223          & 0.235          & 0.230          & 0.240                        & 0.217          & 0.219          & 0.217          & 0.222          & -19841               \\
                                             & PInv   & 0.300          & 0.350          & 0.333          & 0.250          & 0.160          & 0.200          & 0.243          & 0.175          & 0.189          & 0.260          & 0.215          & 0.243          & 0.230          & 0.234          & 0.243                        & 0.227          & 0.226          & 0.223          & 0.228          &  -19931             \\
                                             & Tru    & 0.300          & 0.500          & 0.300          & 0.325          & 0.300          & 0.233          & 0.257          & 0.175          & 0.233          & 0.260          & 0.250          & 0.240          & 0.250          & 0.256          & 0.267                        & 0.246          & 0.249          & 0.250          & 0.251          &   -19639              \\
                                             & Est2   & 0.300          & 0.500          & 0.300          & 0.325          & 0.300          & 0.200          & 0.257          & 0.175          & 0.233          & 0.260          & 0.250          & 0.240          & 0.245          & 0.252          & 0.262                        & 0.243          & 0.246          & 0.248          & 0.248          &     -19634          \\
                                             & Est3   & 0.100          & 0.500          & 0.367          & 0.350          & 0.340          & 0.300          & 0.300          & 0.275          & 0.244          & 0.330          & 0.290          & 0.327          & 0.320          & 0.284          & 0.312                        & 0.306          & 0.299          & 0.302          & 0.306          &  -19526               \\
                                             & Accu   & 0.200          & 0.500          & 0.300          & 0.275          & 0.240          & 0.167          & 0.200          & 0.150          & 0.211          & 0.230          & 0.225          & 0.227          & 0.230          & 0.232          & 0.247                        & 0.231          & 0.230          & 0.231          & 0.232          &   -19576             \\
                                             & CRH    & 0.400          & 0.550          & 0.333          & 0.400          & 0.380          & 0.317          & 0.300          & 0.213          & 0.267          & 0.360          & 0.305          & 0.303          & 0.315          & 0.328          & 0.328                        & 0.317          & 0.314          & 0.318          & 0.318          &       -19703          \\
                                             & SLCA   & \textbf{1.000} & \textbf{1.000} & \textbf{1.000} & \textbf{0.975} & \textbf{0.980} & \textbf{1.000} & \textbf{0.986} & \textbf{0.925} & \textbf{0.944} & \textbf{0.970} & \textbf{0.955} & \textbf{0.963} & \textbf{0.968} & \textbf{0.964} & \textbf{0.967}               & \textbf{0.963} & \textbf{0.956} & \textbf{0.960} & \textbf{0.962} &   \textbf{-14860}            \\
                                             & GLCA   & 0.400          & 0.600          & 0.467          & 0.500          & 0.380          & 0.450          & 0.414          & 0.350          & 0.389          & 0.430          & 0.435          & 0.410          & 0.425          & 0.436          & 0.432                        & 0.417          & 0.414          & 0.413          & 0.417          &   -19686             \\\cline{2-22}
                                             & \textbf{Dist.}  & 15.033         & 9.165          & 7.874          & 3.464          & 3.873          & 4.36           & 7.280          & 6.325          & 3.873          & 6.245          & 2.449          & 5.657          & 2.828          & 2.236          & 1.732                        & 1.000          & 0.000          & 1.000          & 0.000          &  13.928               \\
\multirow{-14}{*}{\textbf{FP}}                        & \textbf{Cos.}   &  0.799              &   0.948           &    0.952            &  0.993              &               0.989 &         0.985     &     0.960          &   0.974             &0.989               &   0.974             & 0.995               &  0.974              &     0.994           &      0.996          &    0.998                          &   0.999             &   1.000             &   0.999             &  1.000              &     0.848            \\\hline\hline
                                             & Voting & \textbf{1.000} & \textbf{1.000} & \textbf{1.000} & \textbf{1.000} & \textbf{1.000} & \textbf{1.000} & \textbf{1.000} & \textbf{1.000} & \textbf{1.000} & \textbf{1.000} & \textbf{1.000} & \textbf{1.000} & \textbf{1.000} & \textbf{1.000} & \textbf{1.000}               & \textbf{1.000} & \textbf{1.000} & \textbf{1.000} & \textbf{1.000} & \textbf{-1978}  \\
                                             & Sums   & \textbf{1.000} & \textbf{1.000} & \textbf{1.000} & \textbf{1.000} & \textbf{1.000} & \textbf{1.000} & \textbf{1.000} & \textbf{1.000} & \textbf{1.000} & \textbf{1.000} & \textbf{1.000} & \textbf{1.000} & \textbf{1.000} & \textbf{1.000} & \textbf{1.000}               & \textbf{1.000} & \textbf{1.000} & \textbf{1.000} & \textbf{1.000} & \textbf{-1978}  \\
                                             & Avg    & \textbf{1.000} & \textbf{1.000} & \textbf{1.000} & \textbf{1.000} & \textbf{1.000} & \textbf{1.000} & \textbf{1.000} & \textbf{1.000} & \textbf{1.000} & \textbf{1.000} & \textbf{1.000} & \textbf{1.000} & \textbf{1.000} & \textbf{1.000} & \textbf{1.000}               & \textbf{1.000} & \textbf{1.000} & \textbf{1.000} & \textbf{1.000} & \textbf{-1978}  \\
                                             & Inv    & 0.500          & 0.350          & 0.500          & 0.650          & 0.380          & 0.400          & 0.571          & 0.550          & 0.578          & 0.490          & 0.490          & 0.450          & 0.458          & 0.460          & 0.478                        & 0.447          & 0.456          & 0.477          & 0.464          & -10458          \\
                                             & PInv   & 0.500          & 0.350          & 0.400          & 0.575          & 0.340          & 0.317          & 0.529          & 0.525          & 0.500          & 0.420          & 0.420          & 0.407          & 0.410          & 0.412          & 0.425                        & 0.397          & 0.409          & 0.422          & 0.412          & -11755          \\
                                             & Tru    & \textbf{1.000} & \textbf{1.000} & \textbf{1.000} & \textbf{1.000} & \textbf{1.000} & \textbf{1.000} & \textbf{1.000} & \textbf{1.000} & \textbf{1.000} & \textbf{1.000} & \textbf{1.000} & \textbf{1.000} & \textbf{1.000} & \textbf{1.000} & \textbf{1.000}               & \textbf{1.000} & \textbf{1.000} & \textbf{1.000} & \textbf{1.000} & \textbf{-1978}  \\
                                             & Est2   & \textbf{1.000} & \textbf{1.000} & \textbf{1.000} & \textbf{1.000} & \textbf{1.000} & \textbf{1.000} & \textbf{1.000} & \textbf{1.000} & \textbf{1.000} & \textbf{1.000} & \textbf{1.000} & \textbf{1.000} & \textbf{1.000} & \textbf{1.000} & \textbf{1.000}               & \textbf{1.000} & \textbf{1.000} & \textbf{1.000} & \textbf{1.000} & \textbf{-1978}  \\
                                             & Est3   & 0.200          & 0.100          & 0.300          & 0.225          & 0.100          & 0.117          & 0.214          & 0.225          & 0.244          & 0.140          & 0.200          & 0.200          & 0.183          & 0.194          & 0.192                        & 0.183          & 0.193          & 0.189          & 0.184          & -11885          \\
                                             & Accu   & \textbf{1.000} & \textbf{1.000} & \textbf{1.000} & \textbf{1.000} & \textbf{1.000} & \textbf{1.000} & \textbf{1.000} & \textbf{1.000} & \textbf{1.000} & \textbf{1.000} & \textbf{1.000} & \textbf{1.000} & \textbf{1.000} & \textbf{1.000} & \textbf{1.000}               & \textbf{1.000} & \textbf{1.000} & \textbf{1.000} & \textbf{1.000} & \textbf{-1978}  \\
                                             & CRH    & \textbf{1.000} & \textbf{1.000} & \textbf{1.000} & \textbf{1.000} & \textbf{1.000} & \textbf{1.000} & \textbf{1.000} & \textbf{1.000} & \textbf{1.000} & \textbf{1.000} & \textbf{1.000} & \textbf{1.000} & \textbf{1.000} & \textbf{1.000} & \textbf{1.000}               & \textbf{1.000} & \textbf{1.000} & \textbf{1.000} & \textbf{1.000} & \textbf{-1978}  \\
                                             & SLCA   & \textbf{1.000} & \textbf{1.000} & \textbf{1.000} & \textbf{1.000} & \textbf{1.000} & \textbf{1.000} & \textbf{1.000} & \textbf{1.000} & \textbf{1.000} & \textbf{1.000} & \textbf{1.000} & \textbf{1.000} & \textbf{1.000} & \textbf{1.000} & \textbf{1.000}               & \textbf{1.000} & \textbf{1.000} & \textbf{1.000} & \textbf{1.000} & \textbf{-1978}  \\
                                             & GLCA   & \textbf{1.000} & \textbf{1.000} & \textbf{1.000} & \textbf{1.000} & \textbf{1.000} & \textbf{1.000} & \textbf{1.000} & \textbf{1.000} & \textbf{1.000} & \textbf{1.000} & \textbf{1.000} & \textbf{1.000} & \textbf{1.000} & \textbf{1.000} & \textbf{1.000}               & \textbf{1.000} & \textbf{1.000} & \textbf{1.000} & \textbf{1.000} & \textbf{-1978}  \\\cline{2-22}
                                             & \textbf{Dist.}  & 1.000          & 1.000          & 0.000          & 0.000          & 0.000          & 0.000          & 0.000          & 0.000          & 0.000          & 0.000          & 0.000          & 0.000          & 0.000          & 0.000          & 0.000                        & 0.000          & 0.000          & 0.000          & 0.000          & 0.000           \\
\multirow{-14}{*}{\textbf{\textbf{FO}}}                        & \textbf{Cos.}   &  0.999              &   0.999             & 1.000               &     1.000           &               1.000 &  1.000              &  1.000              &     1.000           &      1.000          &     1.000           &      1.000          &    1.000            &   1.000             &   1.000             &   1.000                           &    1.000            &      1.000          &   1.000             &       1.000         &     1.000            \\\hline\hline
                                             & Voting & \textbf{0.400} & 0.200          & 0.233          & 0.250          & 0.320          & \textbf{0.350} & 0.300          & 0.288          & 0.267          & 0.340          & 0.340          & 0.313          & 0.290          & 0.306          & 0.283                        & 0.301          & 0.314          & 0.310          & 0.303          &      -19581           \\
                                             & Sums   & \textbf{0.400} & 0.150          & 0.267          & 0.250          & 0.320          & 0.333          & 0.286          & 0.288          & 0.267          & 0.360          & 0.335          & 0.317          & 0.270          & 0.306          & 0.293                        & 0.299          & 0.310          & 0.307          & 0.303          &     -19549           \\
                                             & Avg    & \textbf{0.400} & 0.150          & 0.300          & 0.250          & 0.320          & \textbf{0.350} & 0.343          & 0.263          & 0.278          & 0.350          & 0.315          & 0.323          & 0.300          & 0.310          & 0.293                        & 0.307          & 0.320          & 0.311          & 0.307          &   -19570              \\
                                             & Inv    & \textbf{0.400} & 0.200          & 0.300          & 0.275          & 0.300          & 0.333          & \textbf{0.371} & 0.263          & 0.289          & 0.370          & 0.325          & 0.337          & 0.303          & 0.292          & 0.300                        & 0.313          & 0.321          & 0.319          & 0.312          &    -19800             \\
                                             & PInv   & \textbf{0.400} & 0.200          & 0.267          & \textbf{0.325} & 0.300          & 0.333          & 0.329          & 0.238          & \textbf{0.311} & 0.340          & 0.340          & 0.330          & 0.305          & 0.292          & 0.308                        & \textbf{0.320} & 0.324          & 0.323          & 0.317          &    -19839             \\
                                             & Tru    & \textbf{0.400} & 0.150          & 0.233          & 0.225          & 0.360          & \textbf{0.350} & 0.271          & \textbf{0.300} & 0.267          & 0.350          & 0.315          & 0.323          & 0.293          & 0.310          & 0.293                        & 0.301          & 0.315          & 0.311          & 0.306          &     -19522            \\
                                             & Est2   & \textbf{0.400} & 0.150          & 0.267          & 0.250          & 0.320          & 0.333          & 0.286          & 0.288          & 0.267          & 0.360          & 0.330          & 0.317          & 0.270          & 0.304          & 0.292                        & 0.299          & 0.309          & 0.306          & 0.302          &         -19548       \\
                                             & Est3   & \textbf{0.400} & 0.200          & 0.333          & 0.175          & 0.300          & 0.250          & 0.214          & 0.275          & 0.300          & \textbf{0.380} & 0.330          & \textbf{0.370} & \textbf{0.348} & 0.320          & \textbf{0.327}               & 0.313          & \textbf{0.333} & \textbf{0.332} & \textbf{0.330} &    \textbf{-19356}            \\
                                             & Accu   & \textbf{0.400} & 0.150          & 0.267          & 0.250          & 0.320          & 0.350          & 0.286          & 0.263          & 0.244          & 0.350          & 0.310          & 0.313          & 0.280          & 0.302          & 0.290                        & 0.304          & 0.310          & 0.308          & 0.302          &         -19551     \\
                                             & CRH    & \textbf{0.400} & \textbf{0.250} & 0.233          & 0.200          & 0.320          & \textbf{0.350} & 0.243          & 0.275          & 0.267          & 0.350          & \textbf{0.350} & 0.297          & 0.273          & 0.294          & 0.293                        & 0.303          & 0.308          & 0.306          & 0.299          &        -19496         \\
                                             & SLCA   & \textbf{0.400} & 0.200          & \textbf{0.367} & 0.250          & \textbf{0.380} & 0.300          & 0.271          & 0.275          & 0.244          & 0.370          & \textbf{0.350} & 0.330          & 0.315          & \textbf{0.334} & 0.300                        & 0.304          & 0.313          & 0.317          & 0.313          &     -19376            \\
                                             & GLCA   & \textbf{0.400} & 0.200          & 0.167          & 0.250          & 0.360          & \textbf{0.350} & 0.286          & 0.288          & 0.267          & 0.350          & 0.325          & 0.323          & 0.295          & 0.314          & 0.297                        & 0.304          & 0.309          & 0.309          & 0.305          &      -19531           \\\cline{2-22}
                                             & \textbf{Dist.}  & 21.726         & 14.457         & 14.318         & 17.635         & 19.875         & 23.601         & 16.279         & 19.950         & 12.884         & 14.177         & 16.217         & 4.243          & 4.359          & 13.153         & 7.937                        & 8.718          & 6.000          & 3.464          & 0.000          &   16.733              \\
\multirow{-14}{*}{\textbf{R}}                         & \textbf{Cos.}   &  0.887              &    0.814            &    0.818            &    0.713            &               0.625 &   0.488             &      0.774          &     0.648           &   0.854             &   0.821             &  0.781              &   0.985             &    0.985            &     0.858           &     0.947                       &    0.936            &       0.971         &    0.990            &    1.000            &       0.778          \\\hline\hline
                                             & Voting & 0.000          & 0.000          & 0.267          & 0.175          & 0.120          & 0.167          & 0.186          & 0.175          & 0.167          & 0.190          & 0.145          & 0.153          & 0.155          & 0.150          & 0.145                        & 0.136          & 0.146          & 0.151          & 0.145          &       -19530         \\
                                             & Sums   & 0.000          & 0.000          & 0.267          & 0.175          & 0.120          & 0.167          & 0.186          & 0.175          & 0.167          & 0.190          & 0.145          & 0.153          & 0.155          & 0.150          & 0.145                        & 0.136          & 0.146          & 0.151          & 0.145          &      -19488           \\
                                             & Avg    & 0.000          & 0.000          & 0.267          & 0.175          & 0.120          & 0.167          & 0.186          & 0.175          & 0.167          & 0.190          & 0.145          & 0.153          & 0.155          & 0.150          & 0.145                        & 0.136          & 0.146          & 0.151          & 0.145          &          -19508       \\
                                             & Inv    & 0.000          & 0.000          & 0.367          & 0.325          & 0.160          & 0.233          & 0.257          & 0.250          & 0.211          & 0.240          & 0.205          & 0.203          & 0.228          & 0.210          & 0.203                        & 0.190          & 0.208          & 0.206          & 0.202          &         -20000       \\
                                             & PInv   & 0.000          & 0.000          & 0.400          & 0.350          & 0.160          & 0.250          & 0.286          & 0.275          & 0.256          & 0.250          & 0.235          & 0.217          & 0.260          & 0.238          & 0.223                        & 0.210          & 0.236          & 0.231          & 0.229          &         -20185        \\
                                             & Tru    & 0.000          & 0.000          & 0.267          & 0.175          & 0.120          & 0.167          & 0.186          & 0.175          & 0.167          & 0.190          & 0.145          & 0.153          & 0.155          & 0.150          & 0.145                        & 0.136          & 0.146          & 0.151          & 0.145          &        -19474         \\
                                             & Est2   & 0.000          & 0.000          & 0.267          & 0.175          & 0.120          & 0.167          & 0.186          & 0.175          & 0.167          & 0.190          & 0.145          & 0.153          & 0.155          & 0.15           & 0.145                        & 0.136          & 0.146          & 0.151          & 0.145          &     -19488            \\
                                             & Est3   & \textbf{0.420} & \textbf{0.420} & \textbf{0.420} & \textbf{0.420} & \textbf{0.420} & \textbf{0.420} & \textbf{0.420} & \textbf{0.420} & \textbf{0.420} & \textbf{0.420} & \textbf{0.420} & \textbf{0.420} & \textbf{0.420} & \textbf{0.420} & \textbf{0.420}               & \textbf{0.420} & \textbf{0.420} & \textbf{0.420} & \textbf{0.420} &  \textbf{-16596}           \\
                                             & Accu   & 0.000          & 0.000          & 0.267          & 0.175          & 0.120          & 0.167          & 0.186          & 0.175          & 0.167          & 0.190          & 0.145          & 0.153          & 0.155          & 0.150          & 0.145                        & 0.136          & 0.146          & 0.151          & 0.145          &        -19493         \\
                                             & CRH    & 0.000          & 0.000          & 0.267          & 0.175          & 0.120          & 0.167          & 0.186          & 0.175          & 0.167          & 0.190          & 0.145          & 0.153          & 0.155          & 0.150          & 0.145                        & 0.136          & 0.146          & 0.151          & 0.145          &       -19482          \\
                                             & SLCA   & 0.200          & 0.200          & 0.333          & 0.225          & 0.220          & 0.233          & 0.243          & 0.238          & 0.278          & 0.270          & 0.270          & 0.300          & 0.240          & 0.246          & 0.268                        & 0.250          & 0.261          & 0.268          & 0.257          &      -19362           \\
                                             & GLCA   & 0.000          & 0.000          & 0.300          & 0.200          & 0.120          & 0.183          & 0.186          & 0.188          & 0.178          & 0.200          & 0.160          & 0.163          & 0.168          & 0.162          & 0.155                        & 0.146          & 0.155          & 0.160          & 0.153          &      -19489           \\\cline{2-22}
                                             & \textbf{Dist.}  & 8.246          & 8.246          & 6.164          & 2.449          & 2.828          & 1.732          & 3.606          & 2.449          & 0.000          & 0.000          & 0.000          & 0.000          & 0.000          & 1.414          & 0.000                        & 0.000          & 0.000          & 0.000          & 0.000          &     13.114            \\
\multirow{-14}{*}{\textbf{Exp}}                       & \textbf{Cos.}   &   0.978             &   0.978             &     0.974           &    0.990            &               0.998 &   0.995             &  0.986              &   0.990             &  1.000              &     1.000            &    1.000             &       1.000          &1.000                 &   0.997             &     1.000                          &      1.000           &       1.000          &     1.000            &    1.000             &  0.873 \\\hline             
\end{tabular}
\end{tiny}
\end{table*}

\section{Experiments}
\label{sec:Experiments} 

\subsection{Experimental Setup}
\label{subsec:Experimental_Setup}
\subsubsection{Evaluation Metrics}
\label{subsec:Evaluation_Metrics}
We implemented all the twelve selected truth discovery methods, ground truth-based evaluation approach, and CompTruthHyp, in Python 3.4.0. All experiments were conducted on a 64-bit Windows 10 Pro. PC with an Intel Core i7-5600 processor and 16GB RAM. We ran each truth discovery method $K$ times (for our experiments, we set $K$ as $10$) and used the above introduced five traditional evaluation metrics, including \emph{precision, recall, accuracy, F$_1$ score}, and \emph{specificity}, as well as \emph{confidence} output by CompTruthHyp, to evaluate their average performance. For single-valued scenarios, as the experimental results show that the rankings of different metrics are all equivalent, we discuss the precision of each method as an example. For multi-valued scenarios, we additionally introduce a new metric, namely \emph{average}, to measure the overall performance of all the methods, which is calculated as the average of the precision, recall, accuracy and specificity of each method.

To validate our approach, i.e., CompTruthHyp, we need to show the ranking of \emph{confidence} of twelve selected methods, is closer than the rankings of various evaluation metrics of the methods derived from sparse/low-quality ground truth, to the real ranking of the performance of the methods derived from the complete ground truth. In this paper, we adopt \emph{Cosine similarity} (denoted as \emph{Cos.}) and \emph{Euclidean distance} (denoted as \emph{Dist.}) to measure the distance of two rankings. For Cos., a bigger value means better performance, while for Dist., a smaller value indicates better performance.

\subsubsection{Synthetic Datasets}
\label{subsec:Synthetic_Datasets}
For the single-valued scenarios, we applied the dataset generator introduced in Section~\ref{subsec:Motivation}, which can be configured to simulate a wide spectrum of truth discovery scenarios (except multi-valued scenarios).  In particular, three parameters determine the scale of the generated dataset, including the number of sources ($|S|$), the number of objects ($|O|$), and the number of distinct values per object ($|V_o|$). The other three parameters determine the characteristics of the generated dataset, including source coverage ($cov$), ground truth distribution per source ($GT$), and distinct value distribution per object ($conf$). We fixed the scale parameters by setting $|S| = 50, |O| = 1, 000$, and $|V_o| = 20$. To better simulate the real-world scenarios, we configured both $cov$ and $conf$ as exponential distributions. By tuning $GT$ as all possible settings, including \emph{uniform, Random, Full-Pessimistic, Full-Optimistic, 80-Pessimistic, 80-Optimistic}, and \emph{Exponential}, we obtained eight groups of synthetic datasets (each group contains $10$ datasets): i) \emph{U25} (Uniform 25), each source provides the same number ($25$\%) of true positive claims; ii) \emph{U75} (Uniform 25), each source provides the same number ($75$\%) of true positive claims; iii) \emph{80P} (80-Pessimistic), $80$\% of the sources provide $20$\% true positive claims; $20$\% of the sources provide $80$\% true positive claims. iv) \emph{80O} (80-Optimistic), $80$\% of the sources provide $80$\% true positive claims. $20$\% of the sources provide $20$\% true positive claims; v) \emph{FP} (Full-Pessimistic), $80$\% of the sources provide always false claims and $20$\% of the sources provide always true positive claims; vi) \emph{FO} (Full-Optimistic), $80$\% of the sources provide always true positive claims and $20$\% of the sources provide always false claims. vii) \emph{R} (Random), the number of true positive claims per source is random; viii) \emph{Exp} (Exponential), the number of true positive values provided by the sources is exponentially distributed. All synthetic datasets were generated with the complete ground truth.

\subsubsection{Real-World Datasets}
\label{subsec:RealWorld_Datasets}
Since most existing datasets for categorical truth discovery~\cite{li2012truth,waguih2014truth} are inapplicable for multi-truth scenarios, we refined two real-world datasets for our experiments, where each object may contain multiple true values. 

\vspace{2mm}
\noindent{\bf{Book-Author dataset}}~\cite{yin2008truth} contains $33,971$ book-author records crawled from \emph{www.abebooks.com}. These records are collected from numerous book websites (i.e., sources). Each record represents a store's positive claims on the author list of a book (i.e., objects). We refined the dataset by removing the invalid and duplicated records, and excluding the records with only minor conflicts to make the problem more challenging---otherwise, even a straightforward method could yield competitive results. We finally obtained $13,659$ distinctive claims, $624$ websites providing values about author name(s) of $677$ books, each book has on average $3$ authors. The ground truth provided by the original dataset was utilized, which covers only $7.91$\% of the objects. The manually collected ground truth is sparse yet with high quality.

\vspace{2mm}
\noindent{\bf{Parent-Children dataset}} was prepared by extracting the parent-children relations from the \emph{Biography dataset}~\cite{pasternack2010knowing}. We obtained $227,583$ claims about the names of the children of $2,579$ people (i.e., objects) edited by $54,764$ users (i.e., sources). In the resulting dataset, each person has on average $2.48$ children. We used the latest editing records as the ground truth, which covers all the objects. However, the quality of ground truth collected in this simple way is obviously very poor.

\subsection{Experiments on Synthetic Datasets}
\label{subsec:Synthetic}
In this set of experiments, we aim to compare the confidence ($\mathcal C_m$) and the precision of twelve methods calculated on different coverages of leveraged ground truth,
denoted as P($1$\%) to P($100$\%),
with their real precision calculated on the complete ground truth, denoted as P($100$\%), on eight groups of synthetic datasets with different settings of ground truth distributions. Table~\ref{tab:Synthetic} shows the experimental results. As the results on U25 and U75 show similar features with 80P, we omit to show them in this paper due to the limited space.

We observe that none of the twelve methods constantly outperforms the others in terms of precision ($100$\%), and a ``one-fits-all'' approach does not seem to be achievable. Based on the best performance values (shown in bold), we can see that the best method changed from dataset to dataset. In some cases, an improved method may not even beat its original version as a result of different features of the applied datasets. For example, while in most datasets 2-Estimates performed better than 3-Estimates, it performed worse than 3-Estimates in \emph{FP} and \emph{R}, where most of the claims provided by most sources could be false. This shows that in such cases, the factor that ``hardness of facts'' should be considered to achieve better truth discovery. This instability of truth discovery methods reveals the importance of evaluating the methods. With a better evaluation approach, users could choose the best method for truth discovery more easily and accurately for a given scenario.

\begin{table}[]
\centering
\vspace{-2mm}
\caption{Experimental results for two real-world datasets (multi-valued scenarios).}
\vspace{-2mm}
\label{tab:Real_World}
\begin{tiny}
\begin{tabular}{|l|l|l|l|l|l|l|l|l|}
\hline
\textbf{Dataset}                   & \textbf{Method} & \textbf{Prec.} & \textbf{Rec.} & \textbf{Acc.} & \textbf{Spec.} & \textbf{F$_1$.}     & \textbf{Avg.} & \textbf{$\mathcal C_m$}   \\ \hline
\multirow{14}{*}{\textbf{Book}}     & Voting & 0.749     & 0.712  & 0.576    & 0.022       & 0.730  & 0.515   & -26258 \\
                           & Sums   & 0.851     & 0.685  & 0.651    & \textbf{0.511}       & 0.759  & 0.674   & -23011 \\
                           & AvgLog    & 0.841     & 0.663  & 0.629    & 0.489       & 0.742  & 0.656   & -23477 \\
                           & Inv    & 0.815     & 0.745  & 0.659    & 0.311       & 0.778  & 0.633   & -23860 \\
                           & PInv   & 0.812     & 0.750  & 0.659    & 0.289       & 0.780  & 0.628   & -23435 \\
                           & Tru    & 0.847     & 0.663  & 0.633    & \textbf{0.511}       & 0.744  & 0.664   & -23303 \\
                           & Est2   & \textbf{0.863 }    & \textbf{0.755}  & \textbf{0.707}    & \textbf{0.511}       & \textbf{0.806}  & \textbf{0.709}   & \textbf{-21915} \\
                           & Est3   & 0.828     & 0.734  & 0.664    & 0.378       & 0.778  & 0.651   & -24907 \\
                           & Accu   & \textbf{0.858}     & \textbf{0.788}  & \textbf{0.725}    & 0.467       & \textbf{0.822}  & \textbf{0.709}   &\textbf{ -21390} \\
                           & CRH    & 0.850     & 0.679  & 0.646    &\textbf{0.511}       & 0.755  & 0.672   & -22751 \\
                           & SLCA   & \textbf{0.861}     & \textbf{0.810}  & \textbf{0.742}    & 0.467       & \textbf{0.835}  & \textbf{0.720}   & \textbf{-21670} \\
                           & GLCA   & 0.846     & 0.658  & 0.629    & \textbf{0.511}       & 0.740  & 0.661   & -23243 \\\cline{2-9}
                           & Dist.  & 5.099     & 13.153 & 11.225    & 13.153       & 10.863 & 4.472   & 0.000  \\
                           & Cos    & 0.980     & 0.865  & 0.901    & 0.861       & 0.909  & 0.985   & 1.000  \\ 
                           \hline\hline
\multirow{14}{*}{\textbf{Children}} & Voting &0.919           & 0.901       & 0.845         & 0.462            & 0.910       &   0.782      &    -330234    \\
                           & Sums   &0.938           & 0.927       &   0.883       &  0.585           &  0.933      &        0.833 & -314582       \\
                           & AvgLog    &0.938       &   0.926     &0.882          &   0.581          & 0.932       &        0.832 &   -314124     \\
                           & Inv    & 0.915          & 0.919       & 0.841         &  0.457           &   0.917     &        0.783 &   -331351     \\
                           & PInv   & 0.912          &  0.912      & 0.839         &   0.454          & 0.912       &        0.779 &   -331523     \\
                           & Tru    &  0.938         & 0.926       &  0.881        &   0.581          &0.932        &        0.832 &   -315231     \\
                           & Est2   & \textbf{0.940 }         &\textbf{0.927 }       & \textbf{0.885}      &            \textbf{0.595} &\textbf{0.933}      &\textbf{0.836}         &   \textbf{-309873}     \\
                           & Est3   &  0.905         &  0.889      & 0.822         & 0.366            &   0.897     &        0.746 &   -340031     \\
                           & Accu   &  \textbf{0.941}         & \textbf{0.928}       & \textbf{0.885}       &            \textbf{0.588} & \textbf{0.934}       &  \textbf{0.836 }      &  \textbf{-310314}      \\
                           & CRH    & 0.938          &  0.927      &  0.883        &0.586             & 0.932       &        0.833 &  -313421      \\
                           & SLCA   &  \textbf{0.942}         &\textbf{0.927}        &  \textbf{0.886}        &             \textbf{0.601}& \textbf{0.935 }      &   \textbf{0.839 }     &  \textbf{-302873}      \\
                           & GLCA   &  0.938         &0.924        & 0.876         &   0.578          &  0.931      &        0.829 &  -321098      \\\cline{2-9}
                           & Dist.  &  2.828      &3.742        &   2.000       &  1.000         &    3.162    &        1.414 &  0.000      \\
                           & Cos.   & 0.994          &0.989        & 0.997         &   0.999          &  0.992      &        0.998 &    1.000    \\ \hline
\end{tabular}
\end{tiny}
\end{table}

From the table, we can see that CompTruthHyp can always identify the best method for the given dataset. For \emph{80P}, \emph{80O}, and \emph{FO}, the majority of methods performed better than random guessing with the real precision bigger than $0.500$. For \emph{FO}, the ranking of precision stayed stable with the coverage of the ground truth tuned from $1$\% to $100$\% and was consistent with the ranking of the real precision. The ranking of the confidence of methods output by CompTruthHyp was also equal to the ranking of their real precision, with $Dist.=0.000$, and $Cos.=1.000$. While CompTruthHyp and ground truth-based evaluation approach showed similar performance on this type of datasets in terms of accuracy, our approach did not cost any efforts for ground truth collection. For \emph{80P} and \emph{80O}, when the coverage of the ground truth increased, the \emph{Dist. } decreased until it reached $0.000$ ($70$\% for \emph{80P}, $80$\% for \emph{80O}), the \emph{Cos. } increased until it reached $1.000$ ($70$\% for \emph{80P}, $80$\% for \emph{80O}). The \emph{Dist. } and \emph{Cos. } of the confidence ranking were $1.414$ and $0.998$ for \emph{80P}, which were as good as those of P($40$\%), while for \emph{80O}, the ground truth-based evaluation approach beated our approach only when they got a ground truth with coverage bigger than $70$\%. Moreover, in real-world datasets, the collection of a ground truth with coverage bigger than $10$\% is a rather challenging task.

For \emph{R}, \emph{FP}, and \emph{Exp}, none of the methods was reliable, except SLCA for \emph{FP}. Almost all the methods performed worse than random guessing with a real precision smaller than $0.500$, and the real precision of those methods was similar with each other. For \emph{R}, with the coverage of the ground truth increased, the \emph{Dist. } and \emph{Cos. } of the precision ranking fluctuated. Even when the coverage reached $90$\%, the \emph{Dist. } was $3.464$, which is still not close enough to the real ranking. Though the \emph{Dist. } of the confidence ranking was $16.733$ and the \emph{Cos.} was $0.778$, which are not close to the real ranking, it performed better than the rankings of P($1$\%), P($4$\%), P($5$\%), P($6$\%), P($8$\%) in terms of \emph{Dist. }, and those of P($4$\%), P($5$\%), P($6$\%), P($7$\%), P($8$\%) in terms of \emph{Cos}. In the case of \emph{FP}, our approach can only identify the best method and performed better than the ground truth-based evaluation approach when the coverage of the ground truth was $1$\%. However, in this case, only the best method performed better than random guessing and all the other methods showed very similar bad performance. For \emph{Exp}, where one source always lies and one source always tells the truth for all the objects and the remaining sources range from $1$\% to $99$\% of values they claim is true. None of the methods was reliable and all of them performed similarly bad. Even in this case, our approach can still find out the best method, i.e., 3-Estimates.

\subsection{Experiments on Real-World Datasets}
\label{subsec:Real-World}
In this section, we 
report similar comparative studies with two real-world datasets for multi-valued scenarios. As precision cannot reflect the overall performance of a method with the complete ground truth (as analyzed in Section~\ref{subsec:Observations}), we compared the confidence ranking of the methods with the ranking of all six metrics calculated on the provided ground truth, including precision, recall, accuracy, specificity, F$_1$ score, and average. Table~\ref{tab:Real_World} shows the experimental results, with all the top-three best performance are in bold. These results also validate the observation that no methods constantly outperformed the others. We also observed that the rankings of different metrics differed from one another, which validates our assertion that any one of those metrics can not individually reflect the overall performance of the methods. All methods performed worse on the Book-Author dataset than on the Parent-Children dataset with lower precision, recall, accuracy, and specificity. The possible reasons contain the poorer quality of sources (poorer ground truth distribution), more missing values (i.e., true values that are missed by all the sources), and the smaller scale of the dataset.

For both datasets, our approach can consistently identify the top-three best methods. The confidence ranking is more similar with the ranking of average than the ranking of the other metrics. This validates that confidence reflects the overall performance of the methods. However, for Book-Author dataset, the \emph{Dist.} of the confidence ranking to average was still bigger than $4.000$ and \emph{Cos.} with average was still lower than $0.99$. This is because the ground truth is relatively sparse, so the ranking of average cannot reflect the real performance ranking of the methods. Another reason is that there may be copying relations among sources, which are neglected by all the methods including our approach. Compared with the Book-Author dataset, the confidence ranking was closer to the rankings of all metrics on the Parent-Children dataset. This is because the ground truth covers all the objects and is obtained by collecting all the latest editions regarding the objects. Though the precision of the ground truth is not $1.000$, the quality of sources in this dataset is relatively high. Therefore, the leveraged ground truth is similar with the complete ground truth.

\section{Conclusions}
\label{sec:Conclusion}
In this paper, we focus on the problem of comparing truth discovery methods without using the ground truth, which has been rarely studied by previous research efforts. We first motivate this study by revealing the bias introduced by sparse ground truth in evaluating the truth discovery methods, by conducting experiments on synthetic datasets with different coverages of the ground truth. Then, we propose a general approach, called \emph{CompTruthHyp}, to solve this issue. In particular, we propose two approaches for single-valued and multi-valued scenarios, respectively. 
Given a dataset, we first calculate the precision of each source by the output of each truth discovery method. Then, based on the source precision and the identified truth, we estimate the probability of observations of the given data set, for each method. Finally, the performance of methods is determined by the ranking of the calculated probabilities. 
Experimental studies on both real-world and synthetic datasets demonstrate the effectiveness of our approach.

This paper is our first step towards truth discovery methods comparison without using the ground truth. Our future work will focus on enhancing the approach by considering more complex application scenarios. For example, we are interested in scenarios with complex source relationships such as copying relations and supportive relations. 

\bibliographystyle{ACM-Reference-Format}
\bibliography{reference} 


\begin{thebibliography}{00}


\ifx \showCODEN    \undefined \def \showCODEN     #1{\unskip}     \fi
\ifx \showDOI      \undefined \def \showDOI       #1{{\tt DOI:}\penalty0{#1}\ }
  \fi
\ifx \showISBNx    \undefined \def \showISBNx     #1{\unskip}     \fi
\ifx \showISBNxiii \undefined \def \showISBNxiii  #1{\unskip}     \fi
\ifx \showISSN     \undefined \def \showISSN      #1{\unskip}     \fi
\ifx \showLCCN     \undefined \def \showLCCN      #1{\unskip}     \fi
\ifx \shownote     \undefined \def \shownote      #1{#1}          \fi
\ifx \showarticletitle \undefined \def \showarticletitle #1{#1}   \fi
\ifx \showURL      \undefined \def \showURL       #1{#1}          \fi
\providecommand\bibfield[2]{#2}
\providecommand\bibinfo[2]{#2}
\providecommand\natexlab[1]{#1}
\providecommand\showeprint[2][]{arXiv:#2}

\bibitem[\protect\citeauthoryear{Brin and Page}{Brin and Page}{1998}]%
        {brin1998pagerank}
\bibfield{author}{\bibinfo{person}{Sergey Brin} {and} \bibinfo{person}{Lawrence
  Page}.} \bibinfo{year}{1998}\natexlab{}.
\newblock \showarticletitle{The anatomy of a large-scale hypertextual Web
  search engine}.
\newblock \bibinfo{journal}{{\em Computer Networks and ISDN Systems\/}}
  \bibinfo{volume}{30}, \bibinfo{number}{1--7} (\bibinfo{year}{1998}),
  \bibinfo{pages}{107--117}.
\newblock


\bibitem[\protect\citeauthoryear{Dong, Berti-Equille, Hu, and Srivastava}{Dong
  et~al\mbox{.}}{2010}]%
        {dong2010global}
\bibfield{author}{\bibinfo{person}{Xin~Luna Dong}, \bibinfo{person}{Laure
  Berti-Equille}, \bibinfo{person}{Yifan Hu}, {and} \bibinfo{person}{Divesh
  Srivastava}.} \bibinfo{year}{2010}\natexlab{}.
\newblock \showarticletitle{Global detection of complex copying relationships
  between sources}.
\newblock \bibinfo{journal}{{\em Proc. the VLDB Endowment\/}}
  \bibinfo{volume}{3}, \bibinfo{number}{1-2} (\bibinfo{year}{2010}),
  \bibinfo{pages}{1358--1369}.
\newblock


\bibitem[\protect\citeauthoryear{Dong, Berti-Equille, and Srivastava}{Dong
  et~al\mbox{.}}{2009}]%
        {dong2009integrating}
\bibfield{author}{\bibinfo{person}{Xin~Luna Dong}, \bibinfo{person}{Laure
  Berti-Equille}, {and} \bibinfo{person}{Divesh Srivastava}.}
  \bibinfo{year}{2009}\natexlab{}.
\newblock \showarticletitle{Integrating conflicting data: the role of source
  dependence}.
\newblock \bibinfo{journal}{{\em Proc. the VLDB Endowment\/}}
  \bibinfo{volume}{2}, \bibinfo{number}{1} (\bibinfo{year}{2009}),
  \bibinfo{pages}{550--561}.
\newblock


\bibitem[\protect\citeauthoryear{Dong, Gabrilovich, Heitz, Horn, Lao, Murphy,
  Strohmann, Sun, and Zhang}{Dong et~al\mbox{.}}{2014}]%
        {Dong2014Vault}
\bibfield{author}{\bibinfo{person}{Xin~Luna Dong}, \bibinfo{person}{Evgeniy
  Gabrilovich}, \bibinfo{person}{Geremy Heitz}, \bibinfo{person}{Wilko Horn},
  \bibinfo{person}{Ni Lao}, \bibinfo{person}{Kevin Murphy},
  \bibinfo{person}{Thomas Strohmann}, \bibinfo{person}{Shaohua Sun}, {and}
  \bibinfo{person}{Wei Zhang}.} \bibinfo{year}{2014}\natexlab{}.
\newblock \showarticletitle{Knowledge Vault: A Web-scale approach to
  probabilistic knowledge fusion}. In \bibinfo{booktitle}{{\em Proc. ACM SIGKDD
  international conference on Knowledge discovery and data mining}}.
  \bibinfo{pages}{601--610}.
\newblock


\bibitem[\protect\citeauthoryear{Dong, Saha, and Srivastava}{Dong
  et~al\mbox{.}}{2012}]%
        {dong2012less}
\bibfield{author}{\bibinfo{person}{Xin~Luna Dong}, \bibinfo{person}{Barna
  Saha}, {and} \bibinfo{person}{Divesh Srivastava}.}
  \bibinfo{year}{2012}\natexlab{}.
\newblock \showarticletitle{Less is more: selecting sources wisely for
  integration}.
\newblock \bibinfo{journal}{{\em Proc. the VLDB Endowment\/}}
  \bibinfo{volume}{6}, \bibinfo{number}{2} (\bibinfo{year}{2012}),
  \bibinfo{pages}{37--48}.
\newblock


\bibitem[\protect\citeauthoryear{Fang, Sheng, Wang, and Ngu}{Fang
  et~al\mbox{.}}{2017}]%
        {xiu2017value}
\bibfield{author}{\bibinfo{person}{Xiu~Susie Fang}, \bibinfo{person}{Quan~Z.
  Sheng}, \bibinfo{person}{Xianzhi Wang}, {and} \bibinfo{person}{Anne~H.H.
  Ngu}.} \bibinfo{year}{2017}\natexlab{}.
\newblock \showarticletitle{Value Veracity Estimation for Multi-Truth Ojbects
  via a Graph-Based Approach}. In \bibinfo{booktitle}{{\em Proc. International
  World Wide Web Conference (WWW)}}. \bibinfo{pages}{777--778}.
\newblock


\bibitem[\protect\citeauthoryear{Galland, Abiteboul, Marian, and
  Senellart}{Galland et~al\mbox{.}}{2010}]%
        {galland2010corroborating}
\bibfield{author}{\bibinfo{person}{Alban Galland}, \bibinfo{person}{Serge
  Abiteboul}, \bibinfo{person}{Am{\'e}lie Marian}, {and}
  \bibinfo{person}{Pierre Senellart}.} \bibinfo{year}{2010}\natexlab{}.
\newblock \showarticletitle{Corroborating information from disagreeing views}.
  In \bibinfo{booktitle}{{\em Proc. ACM International Conference on Web Search
  and Data Mining (WSDM)}}. \bibinfo{pages}{131--140}.
\newblock


\bibitem[\protect\citeauthoryear{Gleich, Constantine, Flaxman, and
  Gunawardana}{Gleich et~al\mbox{.}}{2010}]%
        {Gleich2010surfer}
\bibfield{author}{\bibinfo{person}{David~F. Gleich}, \bibinfo{person}{Paul~G.
  Constantine}, \bibinfo{person}{Abraham~D. Flaxman}, {and}
  \bibinfo{person}{Asela Gunawardana}.} \bibinfo{year}{2010}\natexlab{}.
\newblock \showarticletitle{Tracking the random surfer: empirically measured
  teleportation parameters in PageRank}. In \bibinfo{booktitle}{{\em Proc.
  International World Wide Web Conference (WWW)}}. \bibinfo{pages}{381--390}.
\newblock


\bibitem[\protect\citeauthoryear{Kleinberg}{Kleinberg}{1999}]%
        {kleinberg1999authoritative}
\bibfield{author}{\bibinfo{person}{Jon~M Kleinberg}.}
  \bibinfo{year}{1999}\natexlab{}.
\newblock \showarticletitle{Authoritative sources in a hyperlinked
  environment}.
\newblock \bibinfo{journal}{{\it J. ACM}} \bibinfo{volume}{46},
  \bibinfo{number}{5} (\bibinfo{year}{1999}), \bibinfo{pages}{604--632}.
\newblock


\bibitem[\protect\citeauthoryear{Li, Li, Gao, Zhao, Fan, and Han}{Li
  et~al\mbox{.}}{2014}]%
        {li2014resolving}
\bibfield{author}{\bibinfo{person}{Qi Li}, \bibinfo{person}{Yaliang Li},
  \bibinfo{person}{Jing Gao}, \bibinfo{person}{Bo Zhao}, \bibinfo{person}{Wei
  Fan}, {and} \bibinfo{person}{Jiawei Han}.} \bibinfo{year}{2014}\natexlab{}.
\newblock \showarticletitle{Resolving conflicts in heterogeneous data by truth
  discovery and source reliability estimation}. In \bibinfo{booktitle}{{\em
  Proc. ACM SIGMOD International Conference on Management of Data}}.
  \bibinfo{pages}{1187--1198}.
\newblock


\bibitem[\protect\citeauthoryear{Li, Dong, Lyons, Meng, and Srivastava}{Li
  et~al\mbox{.}}{2012}]%
        {li2012truth}
\bibfield{author}{\bibinfo{person}{Xian Li}, \bibinfo{person}{Xin~Luna Dong},
  \bibinfo{person}{Kenneth Lyons}, \bibinfo{person}{Weiyi Meng}, {and}
  \bibinfo{person}{Divesh Srivastava}.} \bibinfo{year}{2012}\natexlab{}.
\newblock \showarticletitle{Truth finding on the deep web: is the problem
  solved?}
\newblock \bibinfo{journal}{{\em Proc. the VLDB Endowment\/}}
  \bibinfo{volume}{6}, \bibinfo{number}{2} (\bibinfo{year}{2012}),
  \bibinfo{pages}{97--108}.
\newblock


\bibitem[\protect\citeauthoryear{Li, Gao, Meng, Li, Su, Zhao, Fan, and Han}{Li
  et~al\mbox{.}}{2015}]%
        {li2015survey}
\bibfield{author}{\bibinfo{person}{Yaliang Li}, \bibinfo{person}{Jing Gao},
  \bibinfo{person}{Chuishi Meng}, \bibinfo{person}{Qi Li}, \bibinfo{person}{Lu
  Su}, \bibinfo{person}{Bo Zhao}, \bibinfo{person}{Wei Fan}, {and}
  \bibinfo{person}{Jiawei Han}.} \bibinfo{year}{2015}\natexlab{}.
\newblock \showarticletitle{A survey on truth discovery}.
\newblock \bibinfo{journal}{{\em ACM SIGKDD Explorations Newsletter\/}}
  \bibinfo{volume}{17}, \bibinfo{number}{2} (\bibinfo{year}{2015}),
  \bibinfo{pages}{1--16}.
\newblock


\bibitem[\protect\citeauthoryear{Pasternack and Roth}{Pasternack and
  Roth}{2010}]%
        {pasternack2010knowing}
\bibfield{author}{\bibinfo{person}{Jeff Pasternack} {and} \bibinfo{person}{Dan
  Roth}.} \bibinfo{year}{2010}\natexlab{}.
\newblock \showarticletitle{Knowing what to believe (when you already know
  something)}. In \bibinfo{booktitle}{{\em Proc. International Conference on
  Computational Linguistics (COLING)}}. \bibinfo{pages}{877--885}.
\newblock


\bibitem[\protect\citeauthoryear{Pasternack and Roth}{Pasternack and
  Roth}{2011}]%
        {pasternack2011making}
\bibfield{author}{\bibinfo{person}{Jeff Pasternack} {and} \bibinfo{person}{Dan
  Roth}.} \bibinfo{year}{2011}\natexlab{}.
\newblock \showarticletitle{Making better informed trust decisions with
  generalized fact-finding}. In \bibinfo{booktitle}{{\em Proc. International
  Joint Conferences on Artificial Intelligence (IJCAI)}}.
  \bibinfo{pages}{2324--2329}.
\newblock


\bibitem[\protect\citeauthoryear{Pasternack and Roth}{Pasternack and
  Roth}{2013}]%
        {pasternack2013latent}
\bibfield{author}{\bibinfo{person}{Jeff Pasternack} {and} \bibinfo{person}{Dan
  Roth}.} \bibinfo{year}{2013}\natexlab{}.
\newblock \showarticletitle{Latent credibility analysis}. In
  \bibinfo{booktitle}{{\em Proc. International World Wide Web Conference
  (WWW)}}. \bibinfo{pages}{1009--1020}.
\newblock


\bibitem[\protect\citeauthoryear{Pochampally, Sarma, Dong, Meliou, and
  Srivastava}{Pochampally et~al\mbox{.}}{2014}]%
        {pochampally2014fusing}
\bibfield{author}{\bibinfo{person}{Ravali Pochampally},
  \bibinfo{person}{Anish~Das Sarma}, \bibinfo{person}{Xin~Luna Dong},
  \bibinfo{person}{Alexandra Meliou}, {and} \bibinfo{person}{Divesh
  Srivastava}.} \bibinfo{year}{2014}\natexlab{}.
\newblock \showarticletitle{Fusing data with correlations}. In
  \bibinfo{booktitle}{{\em Proc. ACM SIGMOD International Conference on
  Management of Data}}. \bibinfo{pages}{433--444}.
\newblock


\bibitem[\protect\citeauthoryear{Popat, Mukherjee, Strötgen, and Weikum}{Popat
  et~al\mbox{.}}{2017}]%
        {Popat2017Where}
\bibfield{author}{\bibinfo{person}{Kashyap Popat}, \bibinfo{person}{Subhabrata
  Mukherjee}, \bibinfo{person}{Jannik Strötgen}, {and}
  \bibinfo{person}{Gerhard Weikum}.} \bibinfo{year}{2017}\natexlab{}.
\newblock \showarticletitle{Where the Truth Lies: Explaining the Credibility of
  Emerging Claims on the Web and Social Media}. In \bibinfo{booktitle}{{\em
  Proc. International World Wide Web Conference (WWW)}}.
  \bibinfo{pages}{1003--1012}.
\newblock


\bibitem[\protect\citeauthoryear{Waguih and Berti-Equille}{Waguih and
  Berti-Equille}{2014}]%
        {waguih2014truth}
\bibfield{author}{\bibinfo{person}{Dalia~Attia Waguih} {and}
  \bibinfo{person}{Laure Berti-Equille}.} \bibinfo{year}{2014}\natexlab{}.
\newblock \showarticletitle{Truth discovery algorithms: an experimental
  evaluation}.
\newblock \bibinfo{journal}{{\em arXiv preprint arXiv:1409.6428\/}}
  (\bibinfo{year}{2014}).
\newblock


\bibitem[\protect\citeauthoryear{Wang, Sheng, Fang, Yao, Xu, and Li}{Wang
  et~al\mbox{.}}{2015}]%
        {wang2015integrated}
\bibfield{author}{\bibinfo{person}{Xianzhi Wang}, \bibinfo{person}{Quan~Z.
  Sheng}, \bibinfo{person}{Xiu~Susie Fang}, \bibinfo{person}{Lina Yao},
  \bibinfo{person}{Xiaofei Xu}, {and} \bibinfo{person}{Xue Li}.}
  \bibinfo{year}{2015}\natexlab{}.
\newblock \showarticletitle{An integrated bayesian approach for effective
  multi-truth discovery}. In \bibinfo{booktitle}{{\em Proc. the 24th ACM
  International Conference on Information and Knowledge Management (CIKM)}}.
  \bibinfo{pages}{493--502}.
\newblock


\bibitem[\protect\citeauthoryear{Wang, Sheng, Yao, Li, Fang, and Xu}{Wang
  et~al\mbox{.}}{2016}]%
        {wang2016implications}
\bibfield{author}{\bibinfo{person}{Xianzhi Wang}, \bibinfo{person}{Quan~Z.
  Sheng}, \bibinfo{person}{Lina Yao}, \bibinfo{person}{Xue Li},
  \bibinfo{person}{Xiu~Susie Fang}, {and} \bibinfo{person}{Xiaofei Xu}.}
  \bibinfo{year}{2016}\natexlab{}.
\newblock \showarticletitle{Truth discovery via exploiting implications from
  multi-source data}. In \bibinfo{booktitle}{{\em Proc. the 25th ACM
  International Conference on Information and Knowledge Management (CIKM)}}.
  \bibinfo{pages}{861--870}.
\newblock


\bibitem[\protect\citeauthoryear{Xiao, Gao, Li, Ma, Su, Feng, and Zhang}{Xiao
  et~al\mbox{.}}{2016}]%
        {Xiao2016Confidence}
\bibfield{author}{\bibinfo{person}{Houping Xiao}, \bibinfo{person}{Jing Gao},
  \bibinfo{person}{Qi Li}, \bibinfo{person}{Fenglong Ma}, \bibinfo{person}{Lu
  Su}, \bibinfo{person}{Yunlong Feng}, {and} \bibinfo{person}{Aidong Zhang}.}
  \bibinfo{year}{2016}\natexlab{}.
\newblock \showarticletitle{Towards Confidence in the Truth: A Bootstrapping
  Based Truth Discovery Approach}. In \bibinfo{booktitle}{{\em Proc. ACM SIGKDD
  international conference on Knowledge discovery and data mining}}.
  \bibinfo{pages}{1935--1944}.
\newblock


\bibitem[\protect\citeauthoryear{Yin, Han, and Yu}{Yin et~al\mbox{.}}{2008}]%
        {yin2008truth}
\bibfield{author}{\bibinfo{person}{Xiaoxin Yin}, \bibinfo{person}{Jiawei Han},
  {and} \bibinfo{person}{Philip~S Yu}.} \bibinfo{year}{2008}\natexlab{}.
\newblock \showarticletitle{Truth discovery with multiple conflicting
  information providers on the web}.
\newblock \bibinfo{journal}{{\em IEEE Transactions on Knowledge and Data
  Engineering (TKDE)\/}} \bibinfo{volume}{20}, \bibinfo{number}{6}
  (\bibinfo{year}{2008}), \bibinfo{pages}{796--808}.
\newblock


\bibitem[\protect\citeauthoryear{Zhao and Han}{Zhao and Han}{2012}]%
        {zhao2012probabilistic}
\bibfield{author}{\bibinfo{person}{Bo Zhao} {and} \bibinfo{person}{Jiawei
  Han}.} \bibinfo{year}{2012}\natexlab{}.
\newblock \showarticletitle{A probabilistic model for estimating real-valued
  truth from conflicting sources}. In \bibinfo{booktitle}{{\em Proc.
  International Workshop on Quality in DataBases (QDB), coheld with VLDB}}.
\newblock


\bibitem[\protect\citeauthoryear{Zhao, Rubinstein, Gemmell, and Han}{Zhao
  et~al\mbox{.}}{2012}]%
        {zhao2012bayesian}
\bibfield{author}{\bibinfo{person}{Bo Zhao}, \bibinfo{person}{Benjamin~IP
  Rubinstein}, \bibinfo{person}{Jim Gemmell}, {and} \bibinfo{person}{Jiawei
  Han}.} \bibinfo{year}{2012}\natexlab{}.
\newblock \showarticletitle{A bayesian approach to discovering truth from
  conflicting sources for data integration}.
\newblock \bibinfo{journal}{{\em Proc. the VLDB Endowment\/}}
  \bibinfo{volume}{5}, \bibinfo{number}{6} (\bibinfo{year}{2012}),
  \bibinfo{pages}{550--561}.
\newblock


\end{thebibliography}

\end{document}